%% file: LcPiPi-draft-CL-FH.tex
\documentclass[prl,twocolumn,showpacs,amsmath,amssymb]{revtex4-1}
\usepackage{subfigure}
\usepackage{graphicx}
\usepackage{graphicx}
\usepackage{array}
\usepackage{xcolor}
\usepackage{multirow}

\usepackage{lineno}
\usepackage{LamcNPi-defs}
\usepackage{ulem}

\usepackage{hyperref}
\hypersetup
{
colorlinks=true,
linkcolor=blue,
filecolor=blue,
urlcolor=blue,
citecolor=black,
}

\begin{document}
\normalsize
\parskip=5pt plus 1pt minus 1pt


\title{First measurements of the absolute branching fraction of $\LamCstarB\to \LamC\pi^+\pi^-$ and upper limit on $\LamCstarA\to \LamC\pi^+\pi^-$}

\input{authorlist_2023-08-14.tex}


\begin{abstract}

The absolute branching fraction of the decay \LamCstarPiPiB{} is
measured for the first time to be $(50.7 \pm 5.0_{\rm{stat.}} \pm
4.9_{\rm{syst.}} )\%$ with 368.48 \ipb{} of \ee{} collision data
collected by the BESIII detector at the center-of-mass energies of
$\sqrt{s} = 4.918$ and $4.950$~GeV. This result is lower than the
naive prediction of 67\%, obtained from isospin symmetry, by more than
$2\sigma$, thereby indicating that the novel mechanism referred to as
the \textit{threshold effect}, proposed for the strong decays of
$\LamCstarA$, also applies to $\LamCstarB$. This measurement is
necessary to obtain the coupling constants for the transitions between
$s$-wave and $p$-wave charmed baryons in heavy hadron chiral
perturbation theory.  In addition, we search for the decay
\LamCstarPiPiA{}. No significant signal is observed, and the upper
limit on its branching fraction is determined to be 80.8\% at the 90\%
confidence level.

\end{abstract}

\maketitle

In recent years, a rich mass spectrum of excited charmed baryons has
been
discovered~\cite{Cheng:2021qpd,PDG:2022,Edwards:1995prl,Albrecht:1997plb,PhysRevD.107.032008}.
Identifying their quantum numbers and understanding their properties
are important to study the dynamics of the light quarks in the
environment of a heavy quark.  The strong decays of charmed baryons
are most conveniently described by heavy hadron chiral perturbation
theory (HHChPT), in which heavy quark symmetry and chiral symmetry are
incorporated~\cite{PhysRevD.75.014006,PhysRevD.92.074014}. The chiral
Lagrangian involves several coupling constants for transitions between
$s$-wave and $p$-wave charmed baryons, referred to as $h_2$ to
$h_{15}$~\cite{PhysRevD.46.1148,PhysRevD.56.5483}. Among these, $h_2$
and $h_8$ can be determined from the strong decays of $\LamCstarA$ and
$\LamCstarB$~\cite{Cheng:2021qpd}. These coupling
constants are critical to describe the charmed baryon spectrum and
make predictions of decays into other charmed baryons. However, so
far, the strong decays of $\LamCstarA$ and $\LamCstarB$ are poorly
known due to the scarcity of experimental data~\cite{PDG:2022}. The
existing determinations of $h_2$ and $h_8$ are based on the measured
decay widths of $\LamCstarA$ and $\LamCstarB$. Since the width of
$\LamCstarB$ is nearly zero~\cite{PDG:2022,PhysRevD.107.032008}, only
the upper limit on $h_8$ is provided.  Precise measurements of the
branching fractions of the strong decays of $\LamCstarA$ and
$\LamCstarB$ are important to determine $h_2$ and $h_8$.

In the quark model, \LamCstarA{} and \LamCstarB{} are the lowest-lying
excited states of \LamC{} having spin-parities of 1/2$^-$ and 3/2$^-$,
respectively, and are the degenerate pair of the $p$-wave
state~\cite{Cheng:2021qpd}.  Currently, all observed
decay modes are measured relative to the dominant hadronic
transitions, either $\LamCstarA\to \LamC\pi^+\pi^-$ or $\LamCstarB\to
\LamC\pi^+\pi^-$~\cite{Edwards:1995prl,Albrecht:1997plb}. However, the
absolute branching fractions of these $\pi^+\pi^-$ transitions have
until now never been measured experimentally.  Assuming isospin
symmetry, the ratio between the branching fractions of $\pi^+\pi^-$
and $\pi^0\pi^0$ transitions is 2:1, which is the basis for the
branching fractions of various strong $\LamCstarA$ and $\LamCstarB$
decays quoted in the Particle Data Group
(PDG)~\cite{PDG:2022}. However, isospin symmetry in these processes
has not been verified by any experimental measurement. In
Ref.~\cite{PhysRevD.67.074033}, a novel mechanism called the
\textit{threshold effect} to take into account the limited transition
phase space in these strong decays is proposed. If this mechanism
applies also here, it would result in a 1:1 relation between
$\pi^+\pi^-$ and $\pi^0\pi^0$ transitions in $\LamCstarA$
decay. Furthermore, this mechanism is sensitive to the coupling
constants~\cite{Cheng:2021qpd,PhysRevD.67.074033}, and
the measurements of their branching fractions are crucial to determine
their coupling constants.

In addition, the internal structure of $\LamCstarA$ and $\LamCstarB$
have received much attention since the discoveries of these two
baryons. Significantly, different decay properties of $\LamCstarA$ and
$\LamCstarB$ are observed in
experiments~\cite{Frabetti:1994prl,Edwards:1995prl,Frabetti:1996plb,Albrecht:1993plb,Albrecht:1997plb,Aaltonen:2011prb}. One
example is the decay width: while being approximate 2.6~MeV in the
case of \LamCstarA{}, it is smaller than 1~MeV for
\LamCstarB{}~\cite{PDG:2022}. In addition, the \LamCstarA{} is located
at the $\Sigma_c \pi$ mass threshold, and predominantly decays through
the intermediate state $\Sigma_{c}$ to the hadronic final states
$\LamC\pi\pi$, where $\Sigma_c$ represents the isospin triplet
$\Sigma_c^0$, $\Sigma_c^+$, and $\Sigma_c^{++}$.  However, the
\LamCstarB{} decays into $\Sigma_c$ are highly suppressed.  Exotic
features, such as a molecule-like state rather than a conventional
three-quark structure, have been proposed as explanations for the
difference~\cite{PhysRevD.95.114018,PhysRevD.92.014036,PhysRevD.101.014018,PhysRevD.101.014018,EPJC.81.224}.
Other interpretations include dynamically generated meson-baryon
states~\cite{PhysRevD.101.014018}, analogous to the case of
$\Lambda(1405)$ and $\Lambda(1520)$~\cite{Moriya:2014prl,
  Kamano:2015prc}, or a state with large pentaquark
components~\cite{EPJC.81.224}.

In this Letter, we report the first measurement of the absolute
branching fraction of $\LamCstarB\to$~$\LamC\pi^+\pi^-$ and the upper
limit on $\LamCstarA$~$\to$~$\LamC\pi^+\pi^-$, obtained from the
processes of $\ee\to\ALamC\LamCstarA$ and $\ALamC\LamCstarB$.
We use the data collected with the BESIII
at center-of-mass (c.m.)~energies of 4.918 and
4.950~GeV~\cite{Baician:2023}. The integrated luminosities of the data
samples at 4.918 and 4.950 GeV are 208.1 and
160.4~pb$^{-1}$~\cite{BESIII:Lumi}, respectively. Throughout this
Letter, unless explicitly stated, charge conjugate modes are implicitly included.

Details about design and performance of the BESIII detector can be
found in Ref.~\cite{Ablikim:2009aa}. Simulated samples are produced
with Geant4-based~\cite{Agostinelli:2002hh} Monte Carlo (MC)
software, which includes a full implementation of the detector geometry
and response~\cite{Kaixuan:2022} of the BESIII detector. The
simulations are used to determine the efficiency of the detector and
the reconstruction, and to estimate the background.  The inclusive MC
sample, which consists of \LCLC{} events, $D_{(s)}$ production, $\psi$
states produced in initial state radiation processes, and continuum
processes $\ee\to q\bar{q}$ ($q=u,d,s$), is generated to estimate the
potential background. Here, all the known decay modes of charmed
hadrons and charmonia are modeled with {\sc
evtgen}~\cite{Lange:2001uf, Ping:2008zz} using branching fractions
taken from the PDG~\cite{PDG:2022}, while the remaining unknown decays
are modeled with {\sc lundcharm}~\cite{Chen:2000tv,PhysRevLett.31.061301}.  
Final-state radiation from charged final-state particles is incorporated using
{\sc photos}~\cite{Richter-Was:1992hxq}. The processes of these hadron
productions 
are referred to as inclusive background hereafter.

To determine the branching fractions, the approach contains two steps. 
The first is the determination of the total yields for $\LamCstarA$ or $\LamCstarB$,
$N_{\mathrm{tag}}$, which follows the same method in the Ref.~\cite{Junhua:2023} by using the productions
$\ee\to\ALamC\LamCstarA$ and $\ALamC\LamCstarB$. Three hadronic decay modes
($pK^{-}\pi^{+}$, $pK_{S}^{0}$, and $\Lambda\pi^+$) are used to reconstruct the $\LamC$
signal, denoted as ``tagged $\LamC$'' hereafter, and the candidates for $\LamCstarA$ and $\LamCstarB$
are studied with the recoiling mass from the tagged $\LamC$. 
The second step is the determination of signal events for $\LamCstarA$ or $\LamCstarB\to \LamC\pi^+\pi^-$,
$N_{\mathrm{sig}}$, by further selecting candidates for $\ALamC$, $\pi^+$ and $\pi^-$ particles.  
Finally, the branching fractions are calculated as:
\begin{equation} \label{eq:method}
  \mathcal{B}=\frac{N_{\mathrm{sig}} \cdot \sum_{i} \mathcal{B}_{\mathrm{tag}}^{i} \epsilon_{\mathrm{tag}}^{i} } { N_{\mathrm{tag}} \cdot \sum_{i}\mathcal{B}_{\mathrm{tag}}^{i} \epsilon_{\mathrm{sig}}^{i} },
\end{equation}
where $i$ represents each reconstruction mode of the tagged $\LamC$, 
and $\mathcal{B}_{\mathrm{tag}}^{i}$ labels their branching fractions. 
The $\epsilon_{\mathrm{tag}}^{i}$ and $\epsilon_{\mathrm{sig}}^{i}$
are the efficiencies of determining the total yields $N_{\mathrm{tag}}$
and signal yields  $N_{\mathrm{sig}}$, respectively.

\begin{figure}[!htp]
    \begin{center}
    \subfigure{\includegraphics[width=0.23\textwidth]{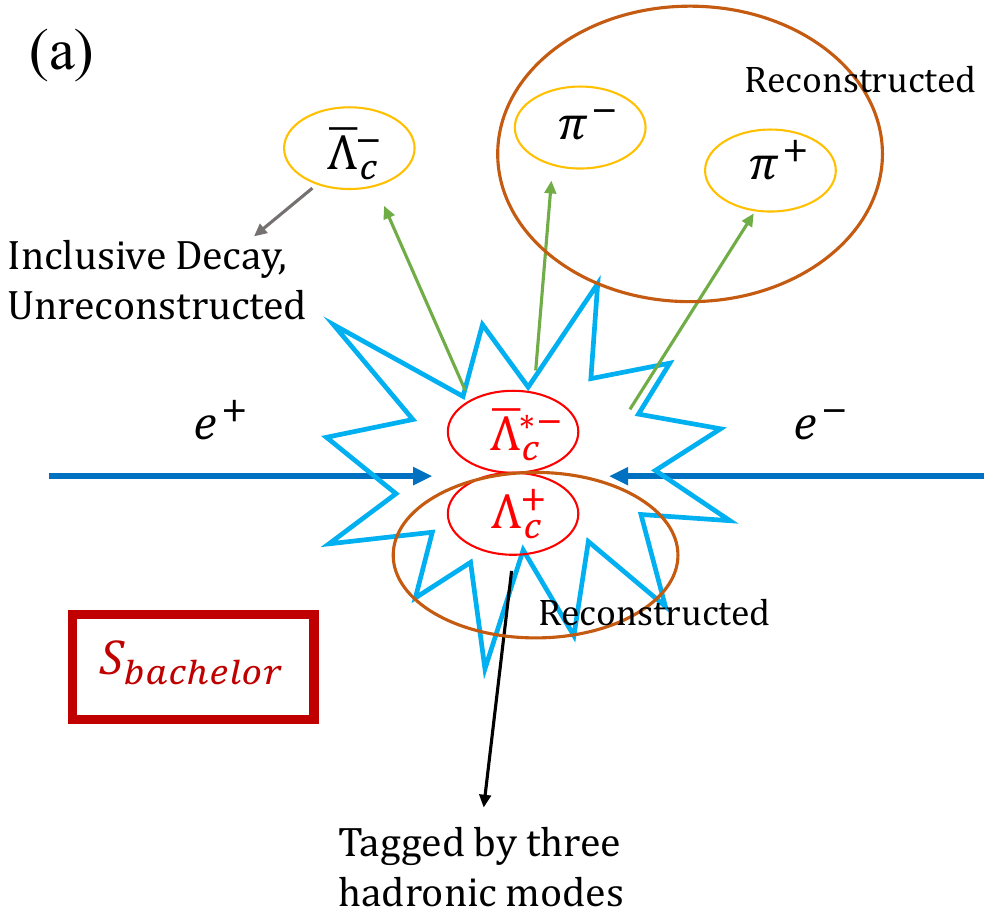}~\label{fig:schematic1}}
    \subfigure{\includegraphics[width=0.23\textwidth]{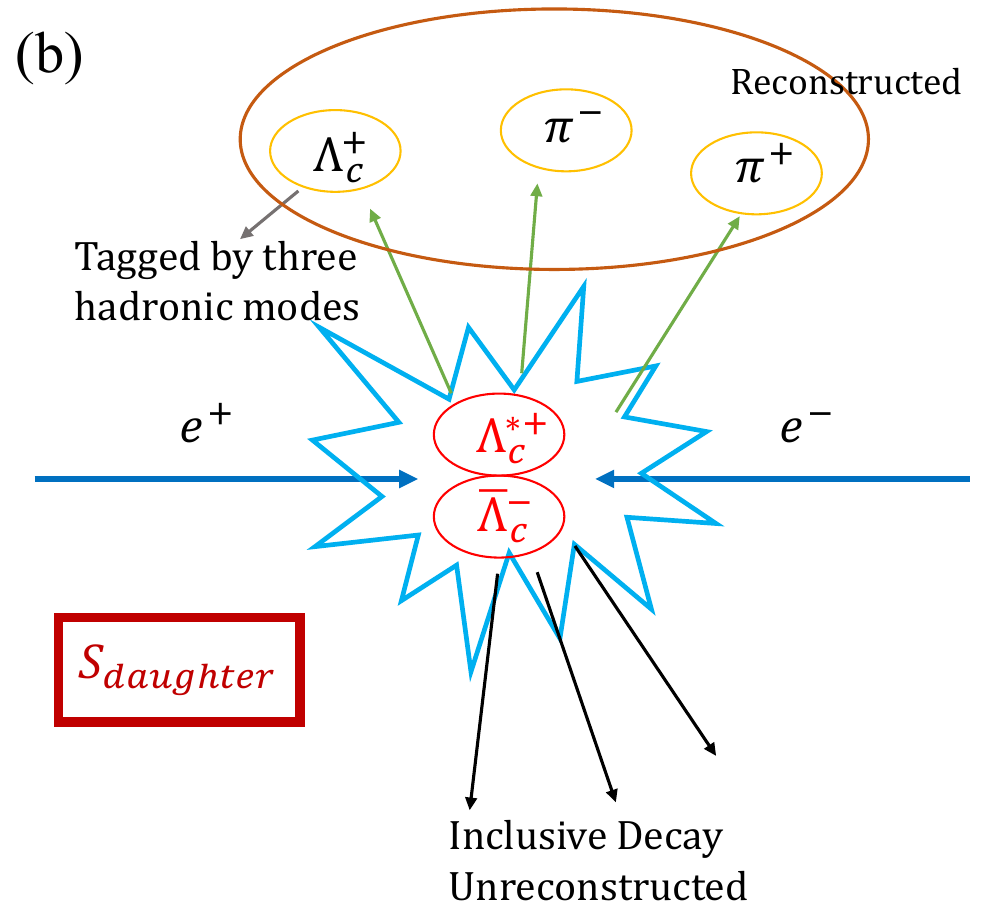}~\label{fig:schematic2}}
     \end{center}
    \vspace{-0.6cm}
    \caption{The signal processes and the partial reconstruction method are schematically presented. 
                (a) The figure corresponds to the $S_{\rm{bachelor}}$ process where the tagged $\LamC$ comes
      directly from the $e^+e^-$ collision and is reconstructed by the three hadronic
      decay modes. (b) This corresponds to the $S_{\rm{daughter}}$ process, where the
      tagged $\LamC$ comes from the decays of the $\Lambda_c^{*+}$, 
      which refers to either $\LamCstarA$ or $\LamCstarB$.}

        \label{schematic}
\end{figure}

To select signal events for $\LamCstarA$ and $\LamCstarB\to \LamC\pi^+\pi^-$, 
a partial reconstruction method is used, 
where the $\LamC$, $\pi^+$ and $\pi^-$ are reconstructed together with another 
unreconstructed $\ALamC$, as demonstrated in Fig.~\ref{schematic}.
The $\LamC$, which decays into the three hadronic modes,
may come from the $e^+ e^-$ collision, as in Fig.~\ref{fig:schematic1}, 
or from the decay of the $\LamCstarA$ or $\LamCstarB$, as in Fig.~\ref{fig:schematic2}. 
The process in Fig.~\ref{fig:schematic1} is referred to as $S_{\rm bachelor}$ and the one
in  Fig.~\ref{fig:schematic2} as $S_{\rm daughter}$. 
The signal MC samples are generated corresponding to the two processes, separately, 
for the two c.m.~energies using the generator {\sc kkmc}~\cite{Jadach:2000ir} 
incorporating initial-state radiation effects and the beam energy spread. 
The $\ALamC$ in both processes is required to decay into any allowed final
states. The line shapes of $\ee\to\ALamC\LamCstarA$ and
$\ALamC\LamCstarB$ 
cross sections in the production for signal MC samples are obtained from the measurements 
by BESIII~\cite{Junhua:2023}. 
In addition, the signal MC samples for the charge-conjugate partners are
also produced for processes $S_{\rm bachelor}$ and $S_{\rm daughter}$, respectively,
where the $\ALamC$ is reconstructed with the three tag modes
$\bar{p}K^{+}\pi^{-}$, $\bar{p}K_{S}^{0}$, and $\bar{\Lambda}\pi^-$,
and the $\LamC$ is required to decay into any allowed final states.

Charged tracks
detected in the helium-based multi-layer chamber (MDC) are required to
be within a polar angle ($\theta$) range of
$\left|\rm{\cos}\theta\right|<0.93$, where $\theta$ is defined with
respect to the $z$-axis, which is the symmetry axis of the MDC. The
distance of closest approach for charged tracks that do not come from
a $\Lambda$ or $K_{S}^{0}$ decay are required to be within $\pm$10 cm
along the $z$-axis and 1 cm in the plane perpendicular to the
beam. Particle identification (PID) is implemented by combining
measurements of the specific ionization energy loss in the MDC
($dE/dx$) and the time-of-flight (TOF) between the interaction point
and the dedicated TOF detector system. Each charged track is assigned
a particle type of pion, kaon or proton, according to which assignment
has the highest probability. For the mode $\LamC\to pK^{-}\pi^{+}$, a
vertex fit is performed to each $p K^-\pi^+$ combination candidate,
and the re-fitted momenta are used in the further study.

Candidates for $K_{S}^{0}$ and $\Lambda$ are reconstructed by their
dominant modes $K_{S}^{0} \to \pi^{+}\pi^{-}$ and $\Lambda \to
p\pi^{-}$, respectively, where the charged tracks are required to have
distances of closest approach to the interaction point that are
within $\pm$20 cm along the $z$-axis. For the $\Lambda$ decay, the PID
requirement is applied to the proton candidate, but not to the
charged pion. A secondary vertex fit is performed to each $K_{S}^{0}$ or
$\Lambda$ candidate, and the re-fitted momenta are used in the further
analysis. A $K_{S}^{0}$/$\Lambda$ candidate requires
the $\chi^2$ of the secondary vertex fit to be less than
100. Furthermore, the decay vertex is required to be separated from
the interaction point by a distance of at least twice the fitted
vertex resolution, and the invariant mass to be within (0.487, 0.511)
GeV/$c^2$ for $\pi^{+}\pi^{-}$ and (1.111, 1.121) GeV/$c^2$ for
$p\pi^{-}$.

\begin{figure}[!htp]
    \begin{center}
    \vspace{-0.3cm}
    \subfigtopskip=0.5pt
    \subfigbottomskip=0.5pt
            \subfigure{\includegraphics[width=0.23\textwidth,height=0.12\textheight, trim=5 0 0 0, clip]{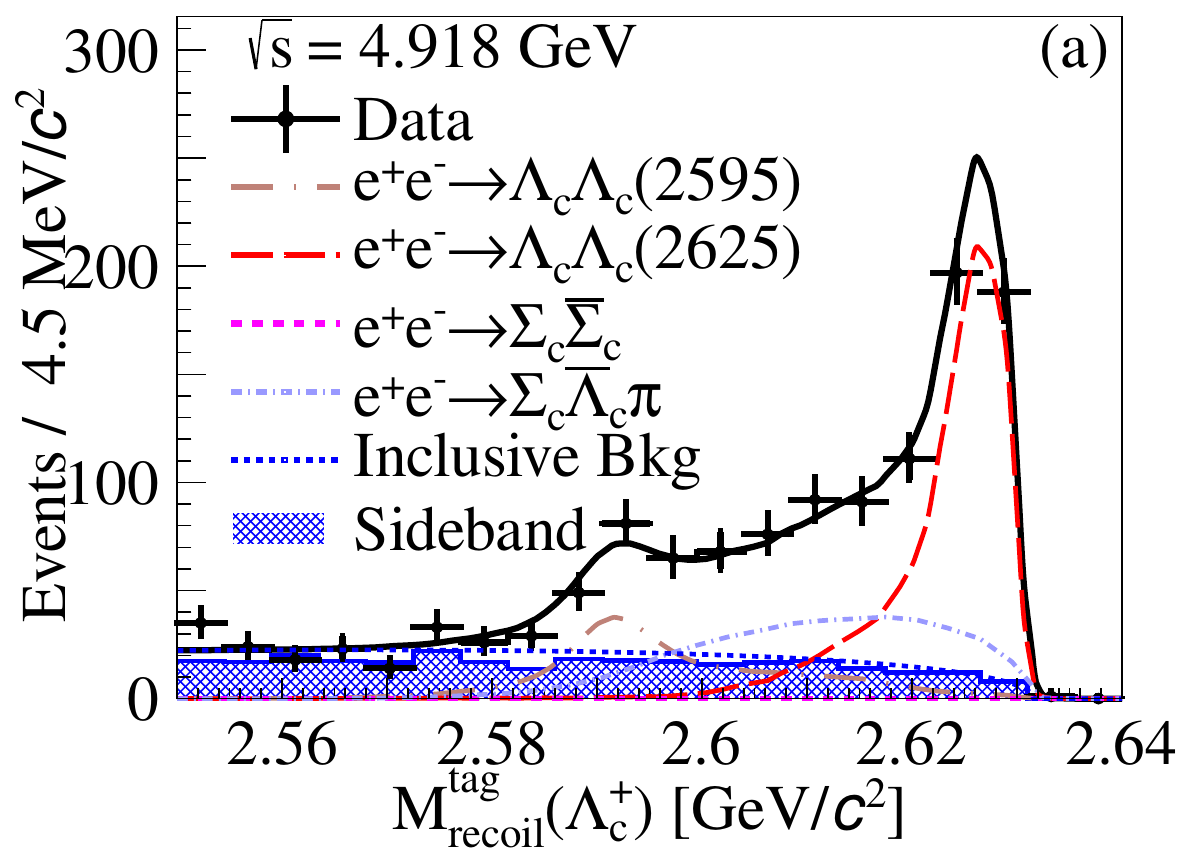}~\label{fig:tag1}} \hspace{-0.34cm}
            \subfigure{\includegraphics[width=0.23\textwidth,height=0.12\textheight, trim=5 0 0 0, clip]{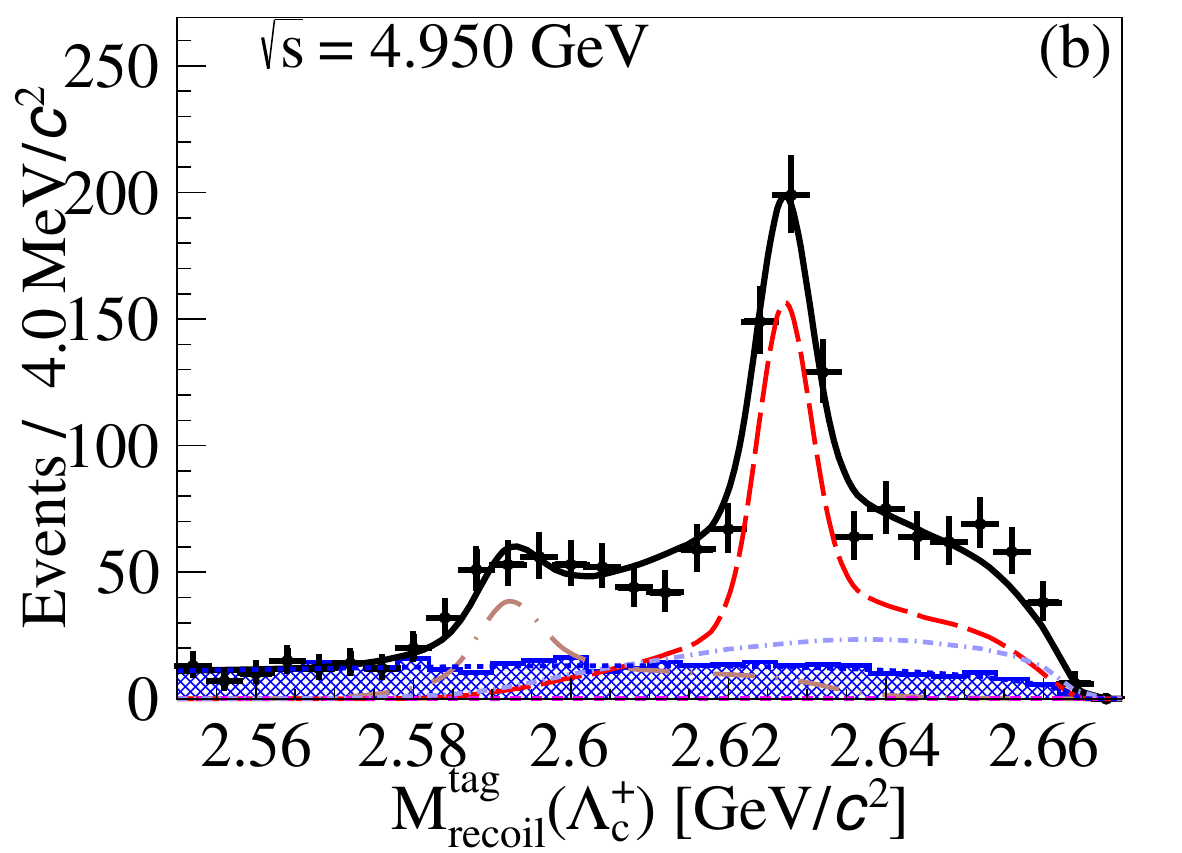}~\label{fig:tag2}} \hspace{-0.34cm}
            \subfigure{\includegraphics[width=0.23\textwidth,height=0.12\textheight, trim=5 0 0 0, clip]{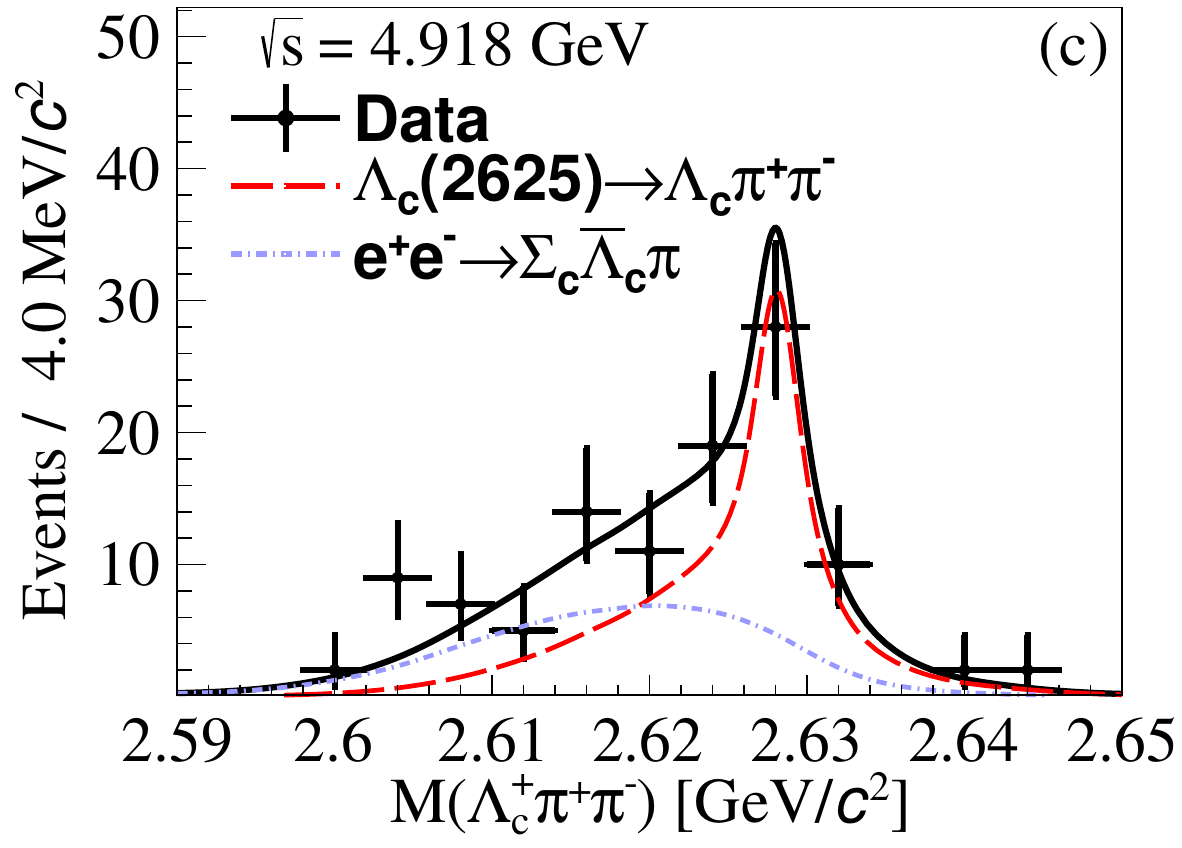}~\label{fig:sig1}} \hspace{-0.34cm}
            \subfigure{\includegraphics[width=0.23\textwidth,height=0.12\textheight, trim=5 0 0 0, clip]{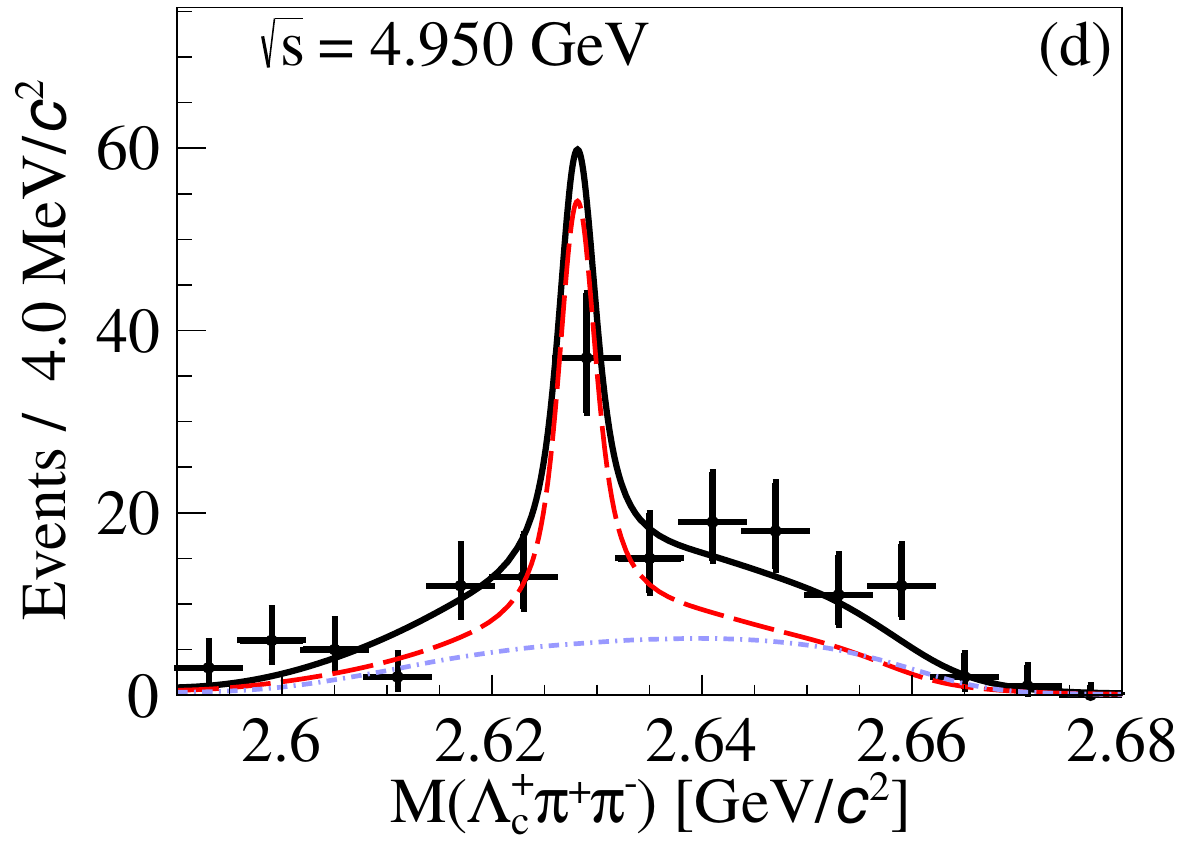}~\label{fig:sig2}} \hspace{-0.34cm}
            \subfigure{\includegraphics[width=0.23\textwidth,height=0.12\textheight, trim=5 0 0 0, clip]{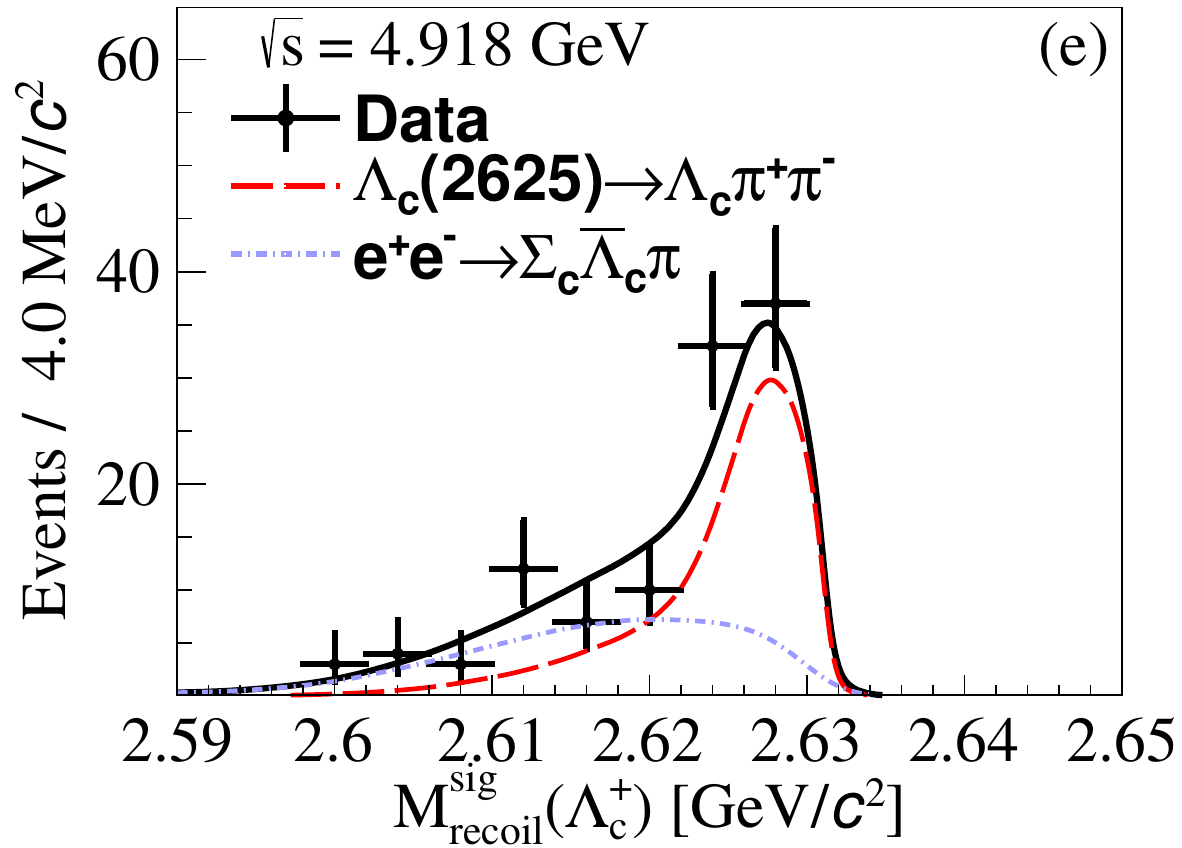}~\label{fig:sig3}} \hspace{-0.34cm}
            \subfigure{\includegraphics[width=0.23\textwidth,height=0.12\textheight, trim=5 0 0 0, clip]{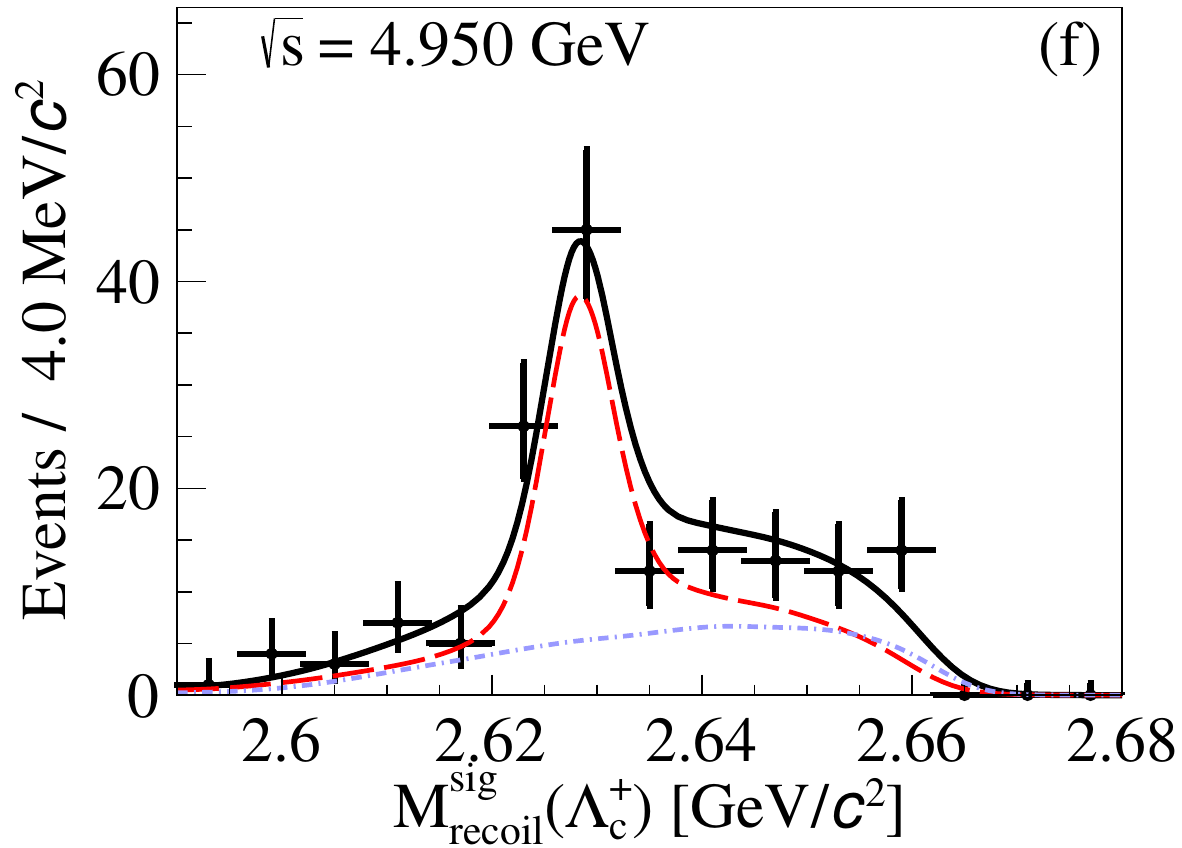}~\label{fig:sig4}} 
	\end{center}
 \vspace{-0.6cm}
    \caption{The distributions of $\MtagrecLc$ at (a) $\sqrt{s} =
      4.918$~GeV  and (b) 4.950~GeV, and 
      distributions of (c, d) $\MLcpipi$, and (e, f) $\MsigrecLc$,
      with all the selection criteria for $\LamC\pi^+\pi^-$.  The
      black points with error bars are data, the solid curves
      represent the fit results, and the dashed ones describe
      individual components including both signal and backgrounds.}
        \label{fig:fit}
\end{figure}

In the first step of determining the total yields $N_{\mathrm{tag}}$, 
all combinations for each decay mode are kept, and their invariant
mass distributions are shown in the supplementary materials~\cite{Supp:2023}.  
The tagged $\LamC$ candidates are required to fall inside the range (2.27,
2.30)~GeV/$c^2$.  The distributions of the recoiling mass from the tagged
$\LamC$, $\MtagrecLc$, are shown in Figs.~\ref{fig:tag1} and
\ref{fig:tag2} by combining the three modes.  There are two components
to the signal at each energy, depending on whether the tagged $\LamC$
originated from the $e^+ e^-$ collision directly or from the decay of
either $\LamCstarA$ or $\LamCstarB$.  If from the $e^+ e^-$ collision
directly, 
narrow resonances $\LamCstarA$ and
$\LamCstarB$ are observed at two energy points $\sqrt{s} =$ 4.918 and
4.950~GeV, from the processes of $\ee\to\LamC\ALamCstarA$ and
$\ee\to\LamC\ALamCstarB$, respectively.  However, if from the decay of
either $\LamCstarA$ or $\LamCstarB$, 
the $\LamCstarA$ and $\LamCstarB$ from the processes $\ee\to\ALamC\LamCstarA$ and
$\ALamC\LamCstarB$ distribute broadly under the resonances.  The
combined signal shapes are displayed in Figs.~\ref{fig:tag1} and \ref{fig:tag2}.

In the second step of determining the $N_{\mathrm{sig}}$, 
in addition to the tagged $\LamC$, a $\pi^+\pi^-$ pair is
selected by imposing the same criteria as for the charged pion in the
mode $\LamC\to pK^{-}\pi^{+}$. A vertex fit is performed to the
$\pi^+$ and $\pi^-$ candidates, and the re-fitted momenta are retained
in the further analysis. In the signal processes, there exists another $\ALamC$
besides the tagged  $\LamC$ and the $\pi^+\pi^-$ pair. 
To improve the detection efficiency, the $\ALamC$ is unreconstructed and 
considered to be a missing particle. If there is more than one combination in
an event, we select only the best combination that gives the minimum
$|\Delta M|$,
\begin{equation} \label{eq:mrec1}
\small
\begin{aligned}
   \Delta M{\rm{ = }}\sqrt {{{\left[ {2 E_{\mathrm{beam}}} - \left(\sum_{i} E_i\right) \right]}^{\rm{2}}} - {{\left( \sum_{i}\vec{p}_i \right)}^{\rm{2}}}}  - m_{{\Lambda _c}}^{\rm PDG},
\end{aligned}
\end{equation}
where $\Ebeam$ is the beam energy and $m_{{\Lambda _c}}^{\rm PDG}$ is
the $\LamC$ nominal mass~\cite{PDG:2022}, $E_i$ and $\vec{p}_i$
represent the energy and momentum, respectively, and $i$ labels the tagged
$\LamC$, $\pi^+$, and $\pi^-$ particles.  To suppress inclusive
background contamination, $\Delta M$ is required to be greater than
$-0.02$~GeV, which keeps more than 97\% of signal.  With all the
selection criteria, the invariant mass distributions of the
$\LamC\pi^+\pi^-$ system, $\MLcpipi$, and the $\LamC$ recoiling mass,
$\MsigrecLc$, are obtained as shown in Figs.~\ref{fig:sig1}, \ref{fig:sig2}, \ref{fig:sig3} and \ref{fig:sig4}.  The resonance $\LamCstarB$ appears in both the
$\MLcpipi$ and the $\MsigrecLc$ distributions at each energy point,
corresponding to the processes $S_{\rm{daughter}}$ and $S_{\rm{bachelor}}$, respectively.  
However, due to quite low detection efficiencies for
the low momentum $\pi^+$ and $\pi^-$, $\LamCstarA$ is not observed.
Here, as for Figs.~\ref{fig:tag1} and \ref{fig:tag2}, the $\MsigrecLc$
signal shapes in Figs.~\ref{fig:sig3} and \ref{fig:sig4} have two
components depending on whether the $\Lambda_c^+$ comes from the $e^+
e^-$ collision or from $\LamCstarB$.  Also the $\MLcpipi$ signal
shapes in Figs.~\ref{fig:sig1} and \ref{fig:sig2} have two components.
There is a narrow component if the $\Lambda_c^+$ is from the decay of
the $\LamCstarB$, and there is a broad component if the $\Lambda_c^+$
is from the $e^+ e^-$ collision and matched with the $\pi$ pair coming
from the $\LamCstarB$. The separated 1D signal shape components are
displayed in the supplementary material~\cite{Supp:2023}, and the combined shapes are
shown in Figs.~\ref{fig:sig1} to \ref{fig:sig4}.

As shown in Figs.~\ref{fig:tag1} and \ref{fig:tag2}, the remaining
contamination of the $\ee\to\ALamC\LamCstarA$ and
$\ALamC\LamCstarB$ candidates is from inclusive background, $\ee \to
\Sigma_{c}\bar{\Sigma}_{c}$ and $\Sigma_{c}\ALamC\pi$.  
The inclusive background 
events are smoothly distributed under the $\LamCstarA$ and
$\LamCstarB$ peaks, and estimated with sideband events $M(\LamC) \in
(2.18, 2.25)$ and $(2.32, 2.39)$~GeV/$c^2$. $\Sigma_{c}$ decays to
$\LamC\pi$ dominantly, but the mass distributions from the processes
$e^+e^- \to \Sigma_{c}\bar{\Sigma}_{c}$ and $\Sigma_{c}\ALamC\pi$ can
be distinguished from those of $\ee\to\LamC\ALamCstarA$ and $\LamC\ALamCstarB$.
In Figs.~\ref{fig:sig1} to \ref{fig:sig4}, the inclusive background contribution is highly suppressed
and is negligible. The remaining contamination comes from $\ee \to
\Sigma_{c}\bar{\Sigma}_{c}$ and $\Sigma_{c}\ALamC\pi$ since they have
the same final states as the signal.

The total yields $N_{\mathrm{tag}}$ for $\LamCstarA$ or $\LamCstarB$ are obtained by performing an unbinned
maximum likelihood fit ($\rm fit_{\rm tag}$) 
to the distributions of $\MtagrecLc$ for each energy point. 
The candidates for $\LamCstarA$ and $\LamCstarB$ are from both $S_{\rm{bachelor}}$ and $S_{\rm{daughter}}$ 
(as Fig.~\ref{schematic}).  
The two contributions have the same
cross section but different detection efficiencies
$\epsilon^i_{\mathrm{tag}}$, and their shapes are obtained with MC simulation.
In these MC samples, the decays of $\LamCstarA$ and $\LamCstarB$ are modeled 
based on the information in the PDG~\cite{PDG:2022}, and both of them decay into 
$\LamC\pi^+\pi^-$ and $\LamC\pi^0\pi^0$ final states, 
where the decay of $\LamCstarA$ via $\Sigma_{c}\pi$ has a rate of 73\%. 
To account for the resolution difference
between data and MC simulation, the narrow $S_{\rm{bachelor}}$ signal
shapes are convolved with Gaussian functions, which are shared between
the two resonances due to the limited sample sizes at individual
energy points. The signal shapes of $S_{\rm{bachelor}}$ and the
broad $S_{\rm{daughter}}$ are merged together in the fit, which
are shown in  Figs.~\ref{fig:tag1}
and \ref{fig:tag2}.  The
inclusive background distributions are modeled by ARGUS
functions~\cite{ARGUS:1990} with the fixed parameters determined by
fitting the sideband events. The magnitudes of the inclusive
background background are free in the $\rm fit_{\rm tag}$.  The
backgrounds from $e^+e^-\to \Sigma_{c}\bar{\Sigma}_{c}$ and
$\Sigma_{c}\ALamC\pi$ are taken into account in the fit, shapes of
which are derived from MC simulations and yields are determined in the
$\rm fit_{\rm tag}$.  The resultant fit curves are depicted in Figs.~\ref{fig:tag1}
and \ref{fig:tag2}.  The significances of the $\LamCstarA$ signal in
the recoil mass distributions from the tagged $\LamC$ are 5.27$\sigma$ and 8.3$\sigma$ at
$\sqrt{s} =$ 4.918 and 4.950 GeV, respectively. The corresponding
values for the $\LamCstarB$ signal 
are 12.7$\sigma$ and 14.0$\sigma$.

The signal yields $N_{\rm sig}$ of $\LamCstarA$ or $\LamCstarB \to \LamC \pi^+\pi^-$ are obtained by
simultaneous two dimensional (2D) unbinned maximum likelihood fits
($\rm fit_{\rm sig}$) 
to the distributions of $\MLcpipi$ and
$\MsigrecLc$ in the $\LamC$ signal regions of the two c.m.~energies,
which have the same branching fraction. 
The 2D signal shapes of $\LamCstarB\to\LamC \pi^+\pi^-$ and those from $e^+e^- \to
\Sigma_{c}\ALamC\pi$ background are modeled by MC simulations, with
the magnitudes free in the $\rm fit_{\rm sig}$. The 2D signal MC distributions are
shown in the supplementary material~\cite{Supp:2023}.  Because the decay $\LamCstarA
\to \LamC \pi^+\pi^-$ is not observed significantly, as shown in
Figs~\ref{fig:sig1} to \ref{fig:sig4}, its contribution is not
considered in the nominal $\rm fit_{\rm sig}$.  
Also, $e^+e^-\to\Sigma_{c}\bar{\Sigma}_{c}$ is not included, since its contribution is
negligible according to the result of ${\rm fit}_{\rm tag}$.  The
resulting fit curves are shown in Figs.~\ref{fig:sig1} to
\ref{fig:sig4}, and the yields are listed in Table~\ref{tab:results}.
The statistical significance of \LamCstarPiPiB{} is $11.9\sigma$, as
calculated with the change of the likelihood values between the fits
with and without the signal component, and accounting for the change
in the number of degrees of freedom.

\begin{table}[!htbp]
  \begin{center}
  \caption{The branching fractions of $\LamCstarA$ and $\LamCstarB\to \LamC\pi^+\pi^-$ and the
  detection efficiencies of $\epsilon_{\rm tag}$ and $\epsilon_{\rm sig}$ for each reconstruction mode at $\sqrt{s}=4.918$ ($4.950$)~GeV, where the efficiencies are expressed in percentage.
  The numbers of events of $N_{\mathrm{tag}}$ and $N_{\mathrm{sig}}$ combine the three reconstruction modes at $\sqrt{s}=4.918$ ($4.950$)~GeV.}
  \renewcommand\arraystretch{1.2}\footnotesize
\begin{tabular}{ l c c c c  }
 \hline\hline
        $\LamC$ decays              &    & $\pkpi$ & $\pks$ & $\Lambda\pi^+$ \\ \hline     
        
         \multirow{4}{2 cm}{$\ALamC\LamCstarB$}  &   $\epsilon_{\mathrm{tag}}$ / \%        & 46.6 (47.4) &  50.0 (49.6) & 38.3 (37.6)   \\ 
         &   $\epsilon_{\mathrm{sig}}$ / \%       & 14.6 (15.0)  &  16.1 (16.2)  & 12.0 (11.9) \\ 
       & $N_{\mathrm{tag}}$ & \multicolumn{3}{c}{418.7 $\pm$ 34.4 (670.9 $\pm$ 55.6)} \\  
      & $N_{\mathrm{sig}}$  & \multicolumn{3}{c}{ 66.8 $\pm$ 6.6 (107.8 $\pm$ 10.6)} \\
      & $\mathcal{B}$ / \% & \multicolumn{3}{c}{$50.7\pm5.0\pm4.9$} \\
 \hline \hline
 
  \multirow{4}{2 cm}{$\ALamC\LamCstarA$}  & $\epsilon_{\mathrm{tag}}$  / \%     &  48.5 (48.8) &  49.9 (49.0) & 38.5 (37.8)  \\   
            & $\epsilon_{\mathrm{sig}}$ / \% & 2.0 (2.5)    &  2.2 (2.7)    & 1.6 (2.1) \\
            &  $N_{\mathrm{tag}}$ & \multicolumn{3}{c}{$135.2 \pm 29.0$ $(210.3 \pm 28.3)$} \\
            &  $N_{\mathrm{sig}}$  & \multicolumn{3}{c}{ $<$ 4.2 (9.0)  } \\ 
            & $\mathcal{B}$ / \% & \multicolumn{3}{c}{ $< 80.8$ } \\
            \hline\hline
 
    \end{tabular}
          \label{tab:results}
  \end{center}
\end{table}

The branching fraction of \LamCstarPiPiB{} is determined to be
$(50.7\pm5.0\pm4.9)\%$, where the first uncertainty is statistical and
the second systematic.  Since no significant \LamCstarA{} signal is
observed in the signal process, we calculate the upper limit of the
branching fraction of \LamCstarPiPiA{} based on the method in
Ref.~\cite{stenson2006exact}. We integrate the likelihood curve as a
function of the branching fraction of
$\LamCstarA$~$\to$~$\LamC\pi^+\pi^-$ from zero to 90\% of the total
area, and the upper limit on its branching fraction at the 90\% confidence
level is $80.8\%$ (see supplementary materials~\cite{Supp:2023}), where both additive and
multiplicative uncertainties are considered.

The systematic uncertainties in the branching fraction measurement are
associated with the total yields $N_{\rm tag}$, the $\pi^{\pm}$
tracking and PID efficiencies, the signal modeling, the requirement of
$\Delta M$, and the fit strategy.  In the measurement of absolute
branching fractions, the selection criteria of the ``tagged $\LamC$''
affect both $N_{\rm tag}$ and $N_{\rm sig}$ in
Eq.~\ref{eq:method}. Therefore, the systematic uncertainties of
detection efficiency and $\mathcal{B}_{\mathrm{tag}}$ cancel.

The uncertainties on $N_{\rm tag}$ for $\LamCstarA$ and $\LamCstarB$,
as listed in Table~\ref{tab:results}, are 11.1\% and 6.6\%,
respectively, which arise from statistical uncertainties in fit$_{\rm
  tag}$.  The uncertainties associated with the $\pi^{\pm}$ tracking
and PID efficiencies are calculated to be 3.5\%, by using the control
sample of $J/\psi\to p \bar p\pi^+\pi^-$~\cite{Weiping:2023}.  The
uncertainty due to the requirement on $\Delta M$ is 0.1\%, which has
been estimated by studies of the resolution difference between data
and MC simulation on the $\Delta M$ distribution.  The uncertainty in
the signal MC modeling is 2.2\%, determined by taking into account
potential $\Sigma_c$ intermediate resonances to the signal MC samples.
The uncertainties due to the fit strategy are 5.4\%, including those
from the wrong match components of $\LamC\pi^+\pi^-$, the modeling of
$e^+e^- \to \Sigma_c\ALamC\pi$ by varying the ratio of production
cross sections of $e^+e^- \to \Sigma_c^0\ALamC\pi^+$,
$\Sigma_c^+\ALamC\pi^0$ and $\Sigma_c^{++}\ALamC\pi^-$, and
consideration of the potential background $e^+e^- \to
\LamC\ALamC\pi^+\pi^-$ by replacing the component $e^+e^- \to
\Sigma_c^0\ALamC\pi^+$ by it in the fit.  All other sources are found
to be negligible.  Assuming all sources are uncorrelated, the total
uncertainties are determined by the quadratic sum of the individual
values, which result in 13.1\% and 9.8\% for the decays $\LamCstarA$
and $\LamCstarB\to \LamC\pi^+\pi^-$, respectively.

In summary, the branching fraction of the strong decay $\LamCstarB
\to \LamC\pi^+\pi^-$ and upper limit for $\LamCstarA\to
\LamC\pi^+\pi^-$ are determined for the first time, in a model
independent approach by using the 368.5 \ipb{} of \ee{} data collected
at $\sqrt s = 4.918$ and $4.950$~GeV with the BESIII detector. The
absolute branching fraction of $\LamCstarB$~$\to$~$\LamC\pi^+\pi^-$ is
measured to be $(50.7 \pm 5.0_{\rm{stat.}} \pm 4.9_{\rm{syst.}})\%$,
which is $2\sigma$ lower than theoretical prediction
67\%~\cite{PDG:2022} assuming isospin symmetry in the decay of
$\LamCstarB$. This indicates the novel mechanism ``threshold effect''
proposed in the decays of $\LamCstarA$~\cite{PhysRevD.67.074033}
potentially also exists in $\LamCstarB$ decays. Our result provides
critical experimental input to determine the coupling constants in the
HHChPT~\cite{PhysRevD.75.014006,PhysRevD.92.074014}.  In addition, the
measured absolute branching fraction is also essential to calibrate
the relative measurements and guide the search for unknown decays of
\LamCstarB{}.  No discernible signal of the decay \LamCstarPiPiA{} is
observed and the upper limit on its branching fraction at the 90\%
confidence level is 80.8\%.

The BESIII Collaboration thanks the staff of BEPCII and the IHEP computing center for their strong support. This work is supported in part by National Key R\&D Program of China under Contracts Nos. 2020YFA0406300, 2020YFA0406400; National Natural Science Foundation of China (NSFC) under Contracts Nos. 11635010, 11735014, 11835012, 11935015, 11935016, 11935018, 11961141012, 12025502, 12035009, 12035013, 12061131003, 12192260, 12192261, 12192262, 12192263, 12192264, 12192265, 12221005, 12225509, 12235017, 12005311; the Chinese Academy of Sciences (CAS) Large-Scale Scientific Facility Program; the CAS Center for Excellence in Particle Physics (CCEPP); Joint Large-Scale Scientific Facility Funds of the NSFC and CAS under Contract No. U1832207; CAS Key Research Program of Frontier Sciences under Contracts Nos. QYZDJ-SSW-SLH003, QYZDJ-SSW-SLH040; 100 Talents Program of CAS; The Institute of Nuclear and Particle Physics (INPAC) and Shanghai Key Laboratory for Particle Physics and Cosmology; European Union's Horizon 2020 research and innovation programme under Marie Sklodowska-Curie grant agreement under Contract No. 894790; German Research Foundation DFG under Contracts Nos. 455635585, Collaborative Research Center CRC 1044, FOR5327, GRK 2149; Istituto Nazionale di Fisica Nucleare, Italy; Ministry of Development of Turkey under Contract No. DPT2006K-120470; National Research Foundation of Korea under Contract No. NRF-2022R1A2C1092335; National Science and Technology fund of Mongolia; National Science Research and Innovation Fund (NSRF) via the Program Management Unit for Human Resources \& Institutional Development, Research and Innovation of Thailand under Contract No. B16F640076; Polish National Science Centre under Contract No. 2019/35/O/ST2/02907; The Swedish Research Council; The Knut and Alice Wallenberg Foundation, Sweden; U. S. Department of Energy under Contract No. DE-FG02-05ER41374


\end{document}


\normalsize
\parskip=5pt plus 1pt minus 1pt

\title{ Supplementary materials of First measurement of absolute branching fractions of $\LamCstarA$ and $\LamCstarB\to \LamC\pi^+\pi^-$ }



\maketitle
\begin{figure}[!htp]
    \begin{center}
            \subfigure{\includegraphics[width=0.42\textwidth]{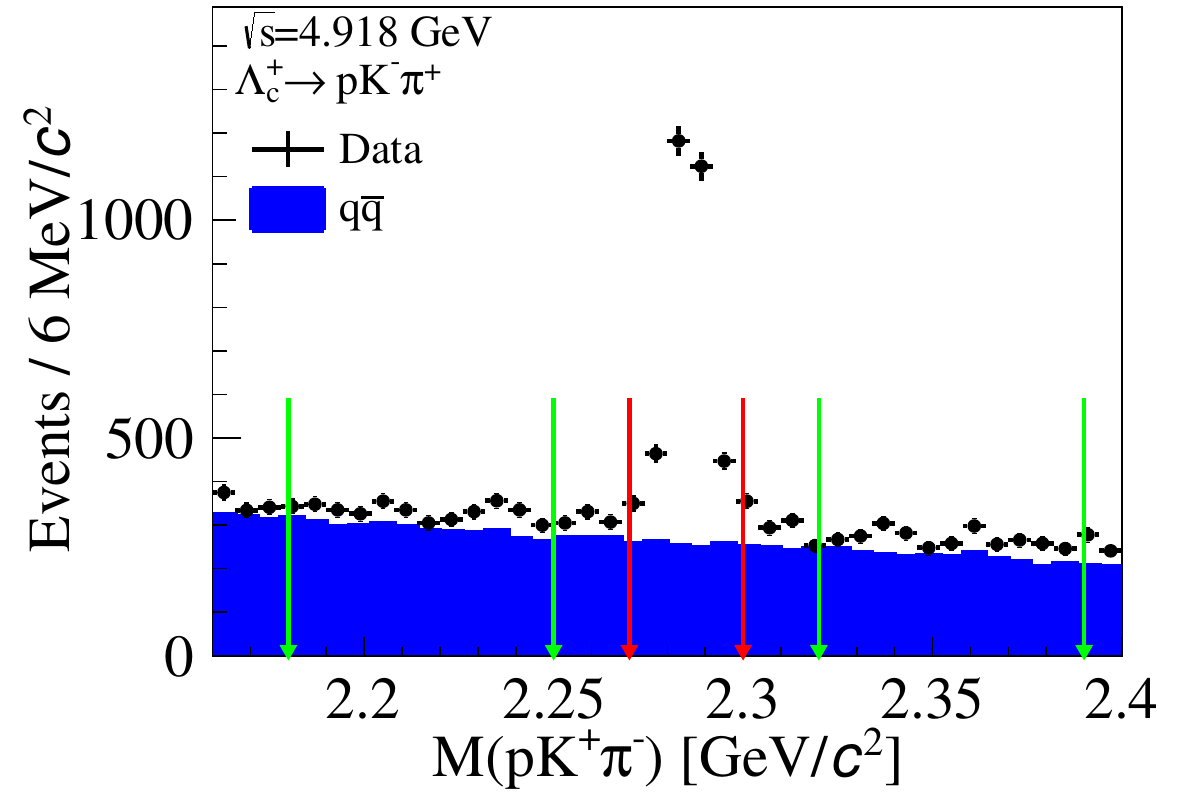}~\label{fig:4945-1}} 
            \subfigure{\includegraphics[width=0.42\textwidth]{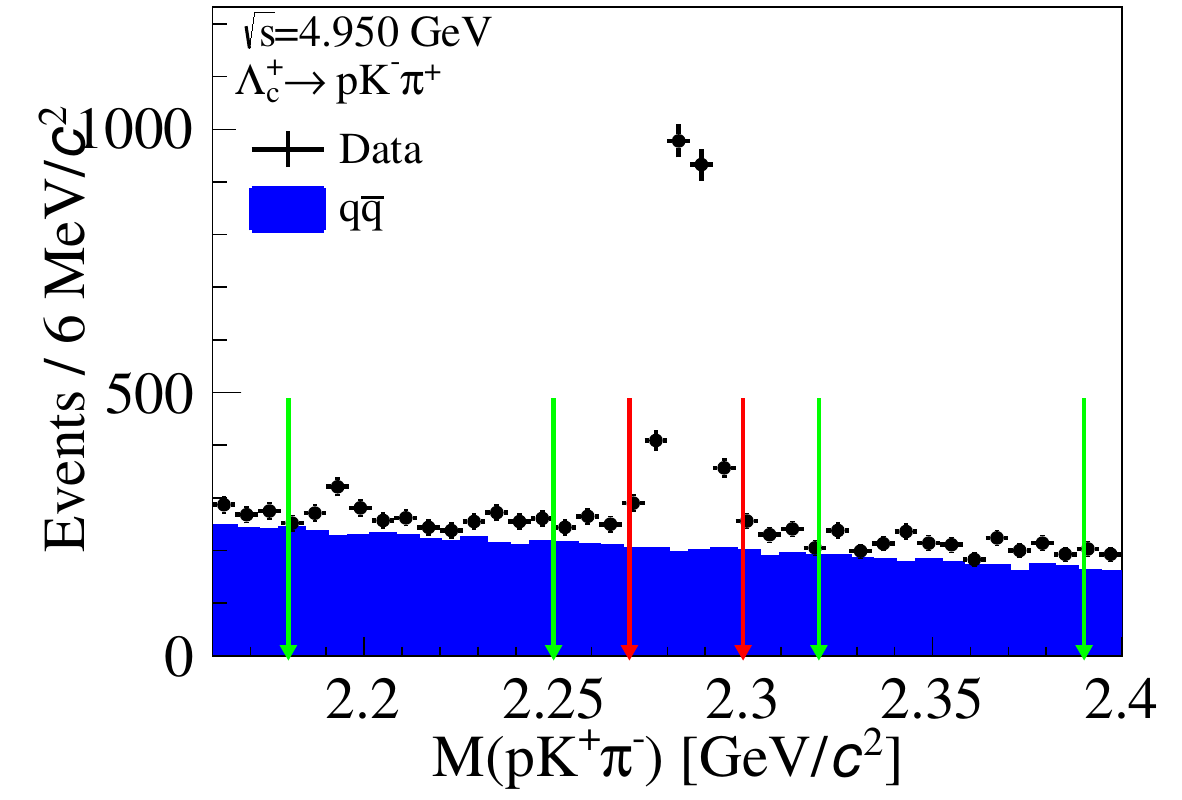}~\label{fig:4945-2}}  
            
            \subfigure{\includegraphics[width=0.42\textwidth]{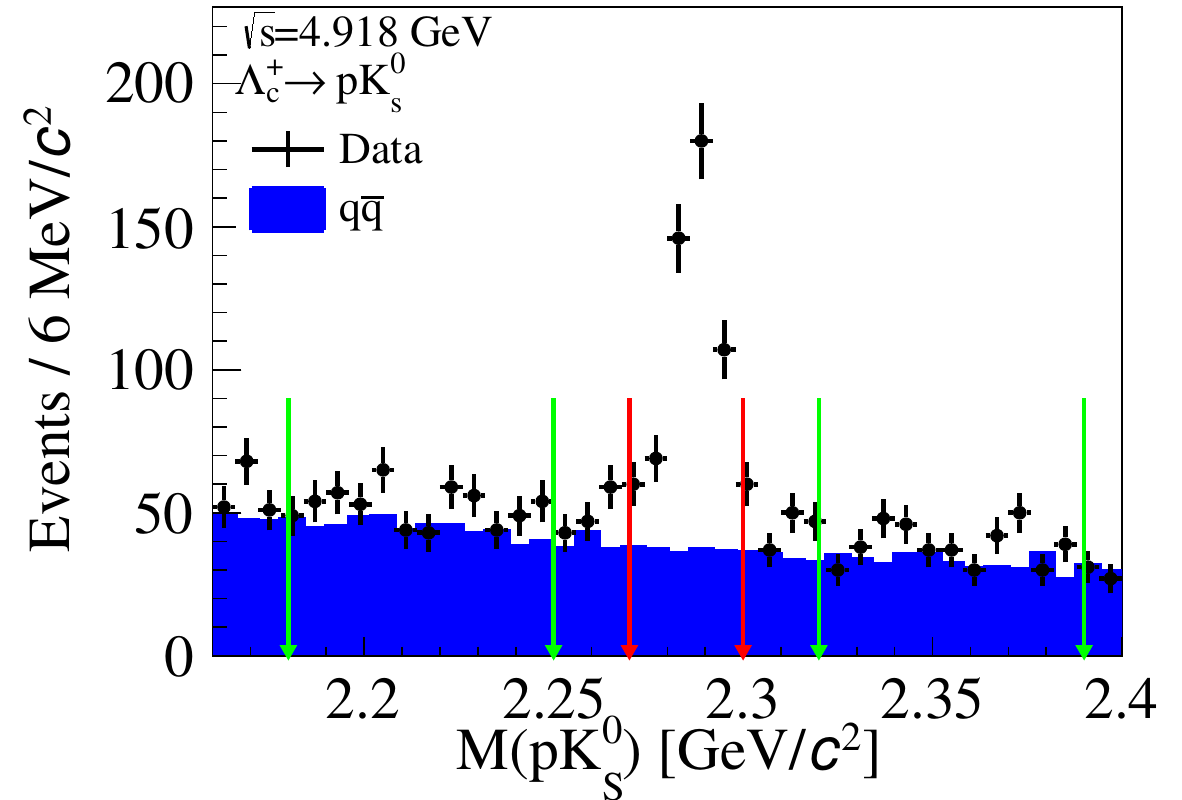}~\label{fig:4945-3}}
            \subfigure{\includegraphics[width=0.42\textwidth]{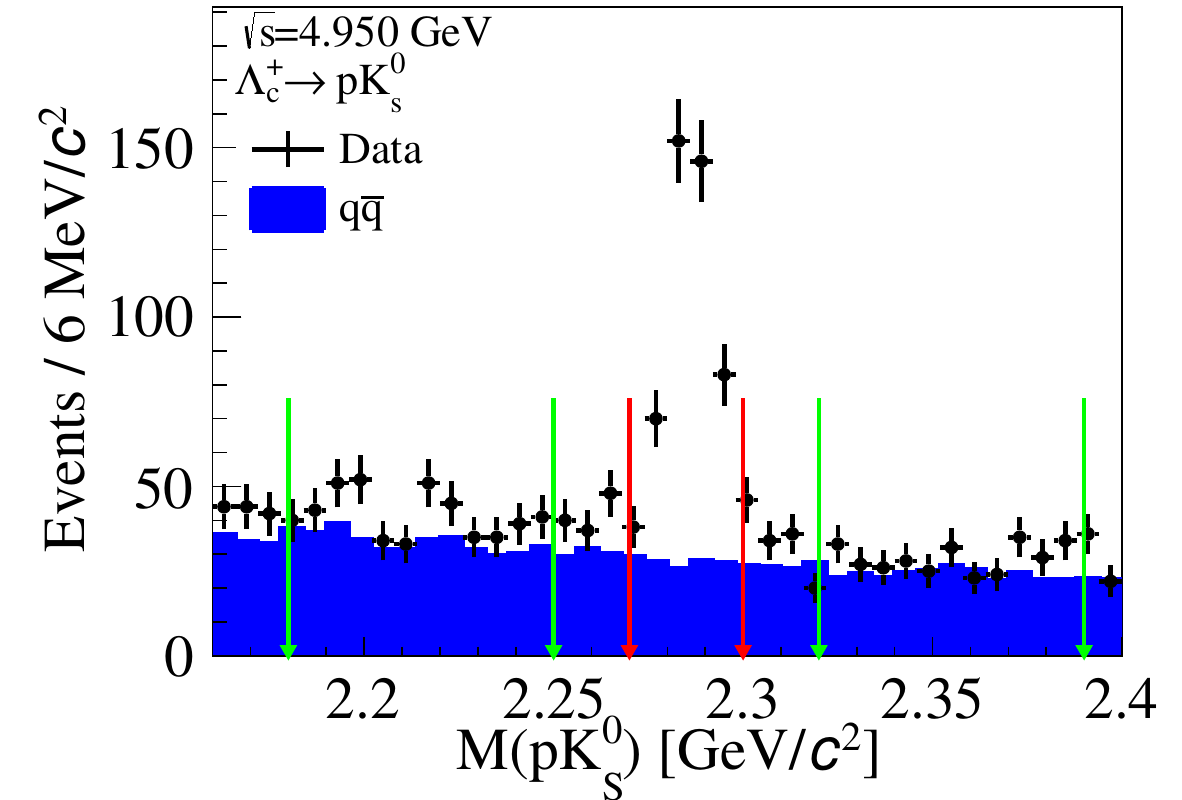}~\label{fig:4946-1}}
            
            \subfigure{\includegraphics[width=0.42\textwidth]{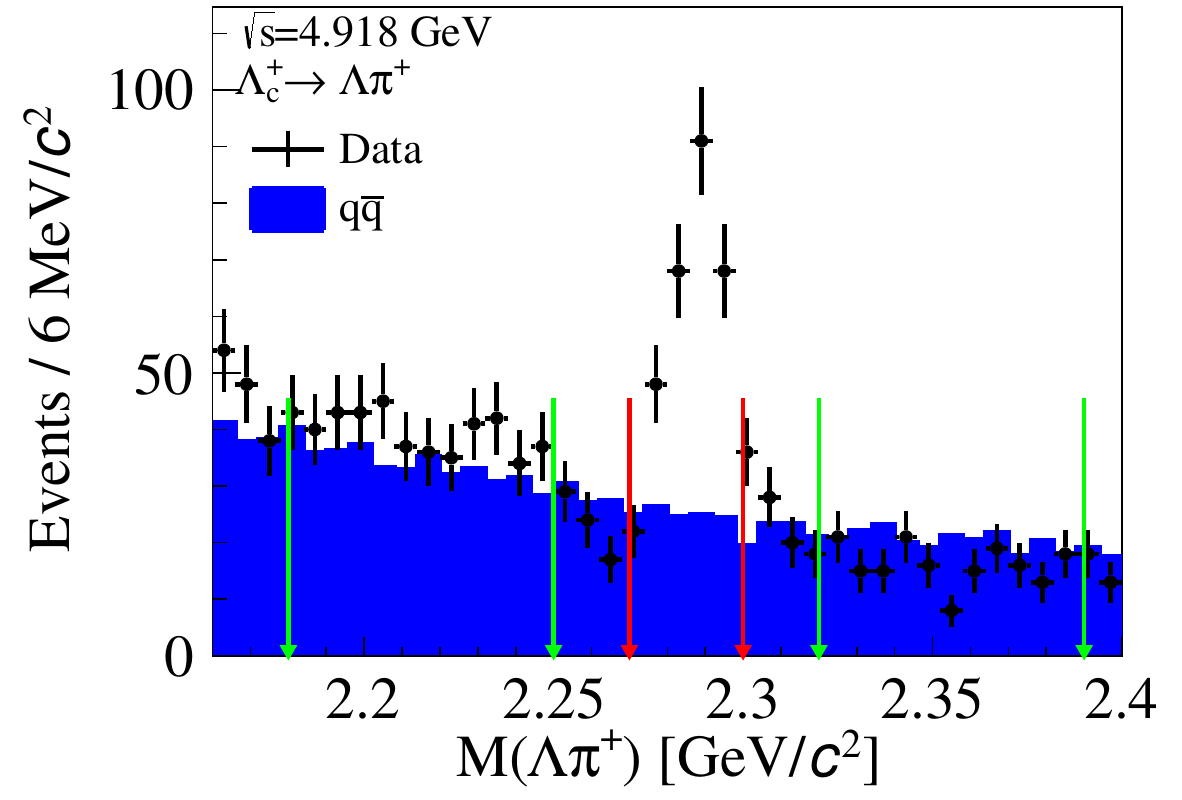}~\label{fig:4946-2}}
            \subfigure{\includegraphics[width=0.42\textwidth]{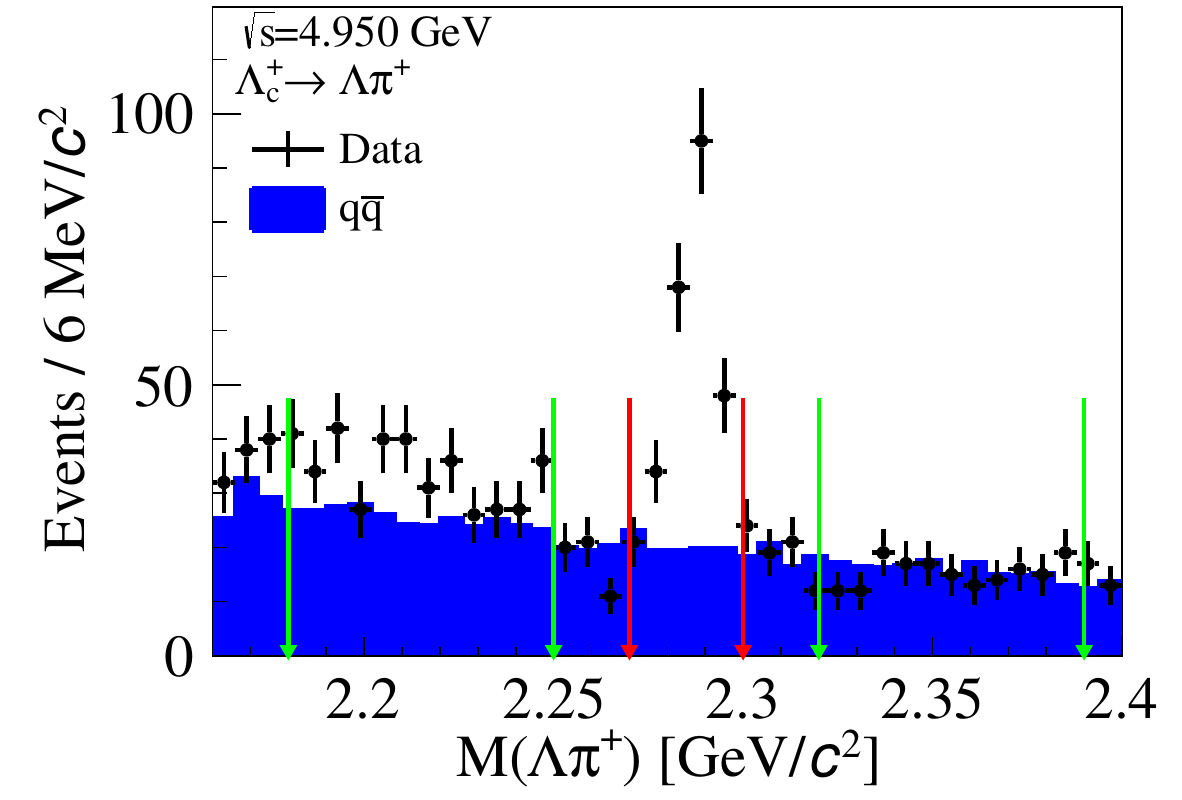}~\label{fig:4946-3}}
	\end{center}
    \caption{The distributions of invariant masses of the tagged $\LamC$ candidates for $\sqrt{s} = 4.918$ (left) and 4.950~GeV (right).
    The red arrows indicate
the signal region (2.270, 2.310) GeV/$c^2$ while the green arrows indicate the sideband regions
(2.180,2.250)~\&~(2.320, 2.390) GeV/$c^2$. }
        \label{fig:mLamC}
\end{figure}
\begin{figure}[!htp]
    \begin{center}
            \subfigure{\includegraphics[width=0.42\textwidth]{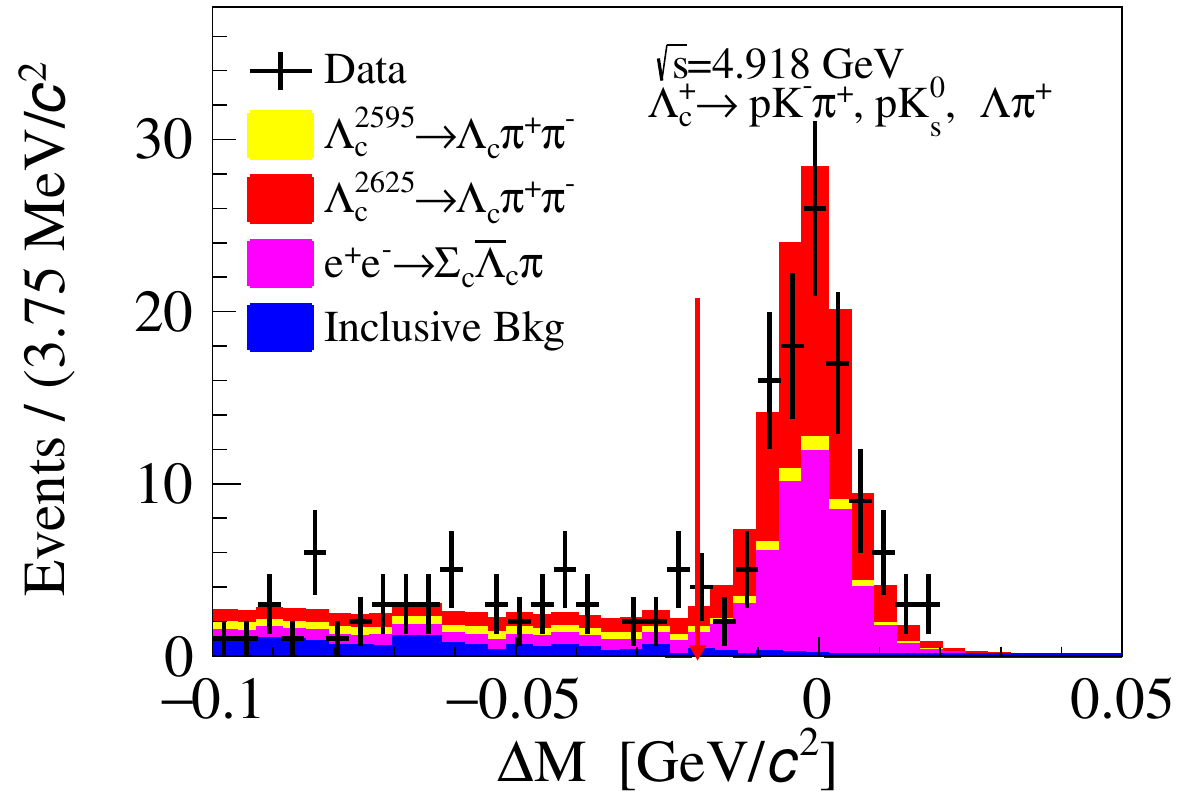}~\label{fig:dM-0}} 
            \subfigure{\includegraphics[width=0.42\textwidth]{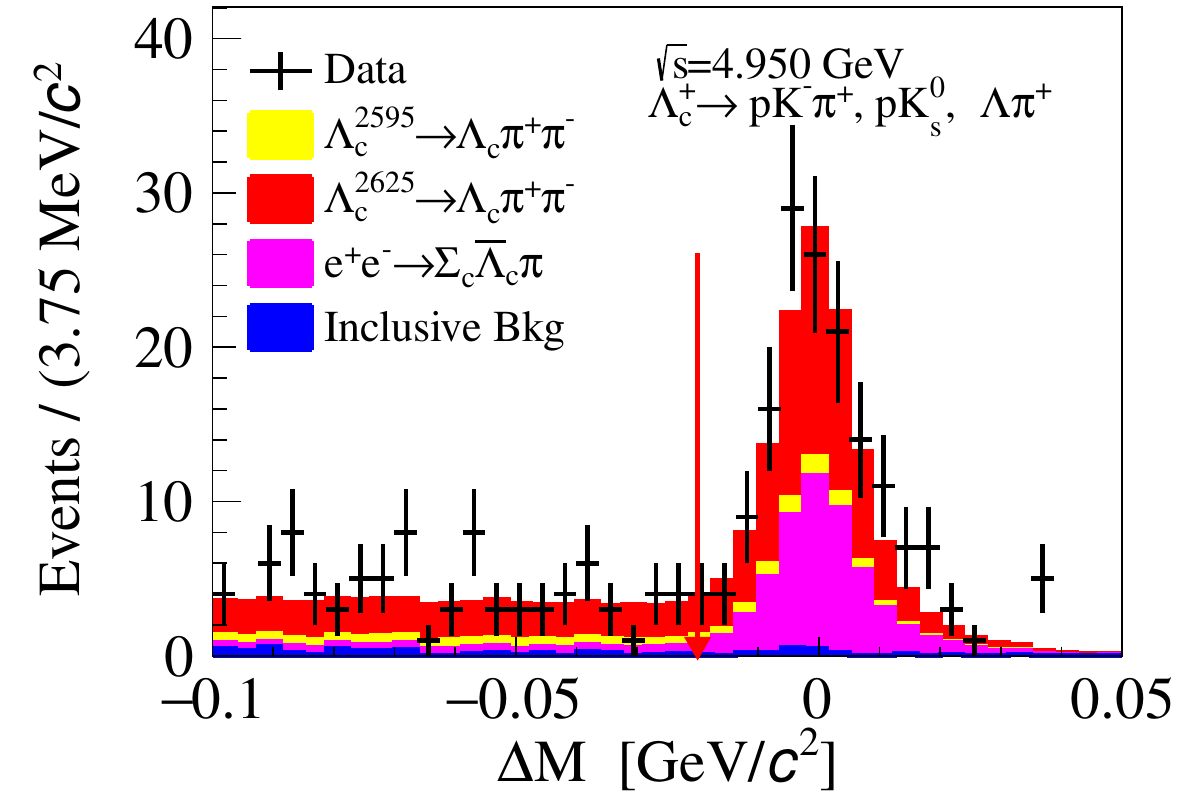}~\label{fig:dM-1}}  
	\end{center}
    \caption{The distributions of $\Delta M$ for $\sqrt{s} = 4.918$ (left) and 4.950~GeV (right).
    The red arrows indicate
the cut $\Delta M > -0.02$ GeV/$c^2$. }
        \label{fig:dM}
\end{figure}
\begin{figure}[!htp]
    \begin{center}
            \subfigure{\includegraphics[width=0.42\textwidth]{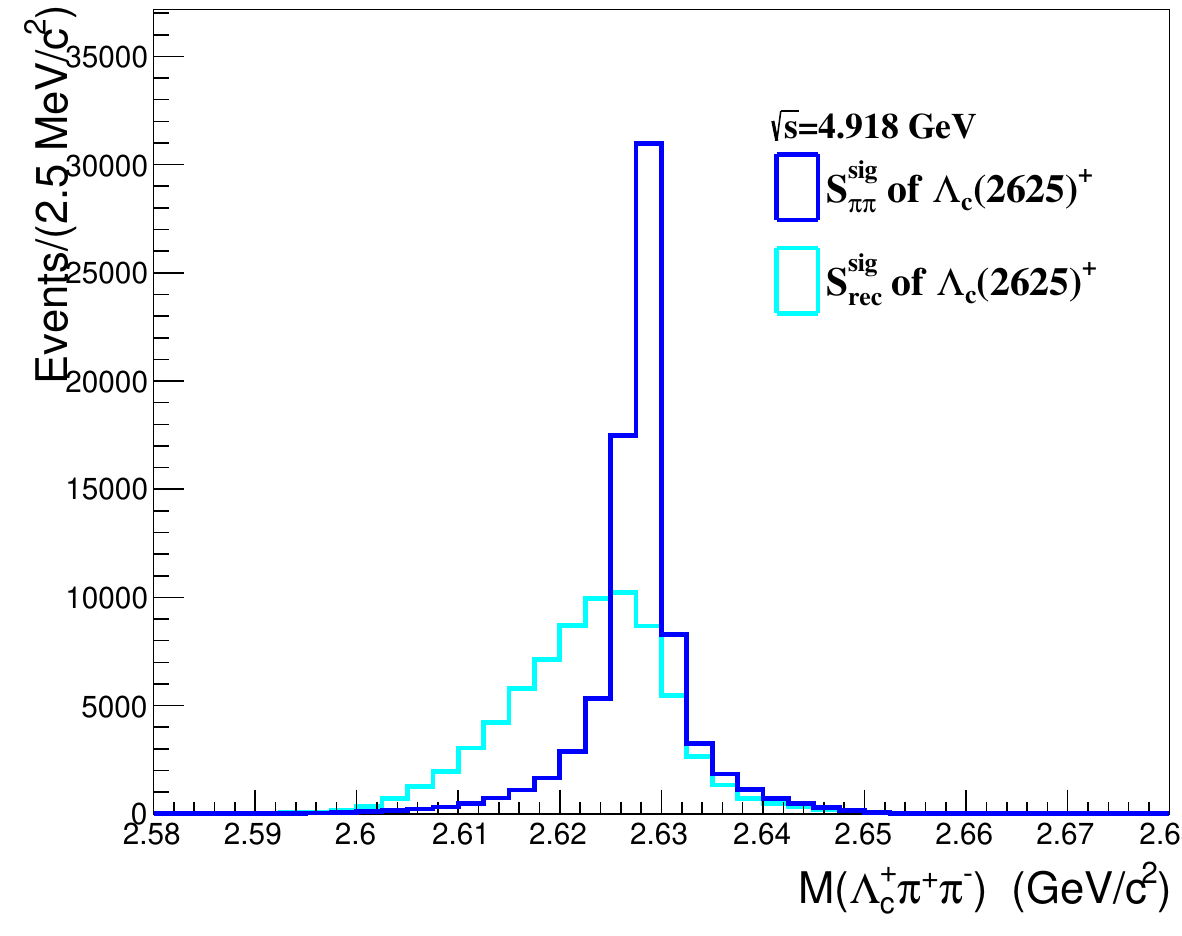}~\label{fig:MLcpipi-4915}} 
            \subfigure{\includegraphics[width=0.42\textwidth]{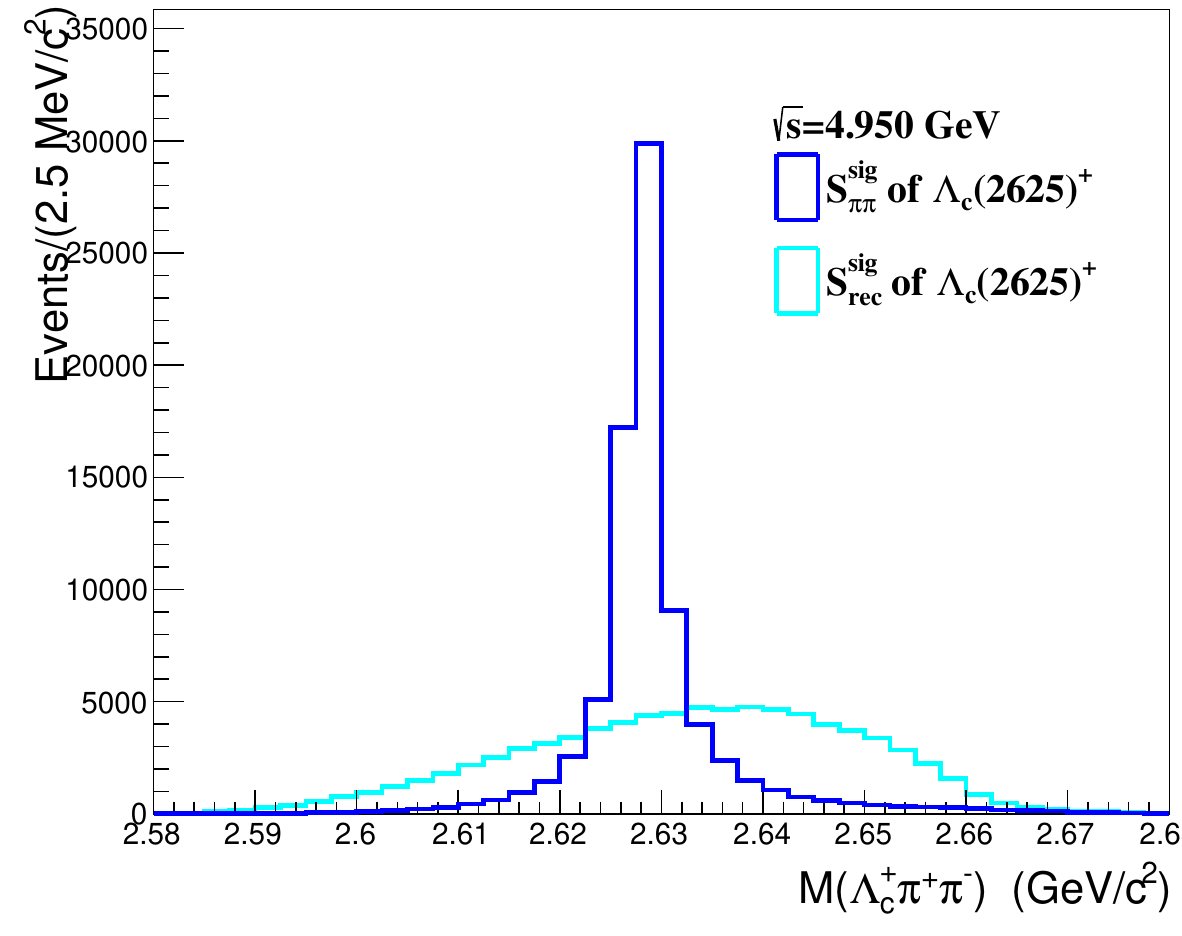}~\label{fig:MLcpipi-4946}}  
            
            \subfigure{\includegraphics[width=0.42\textwidth]{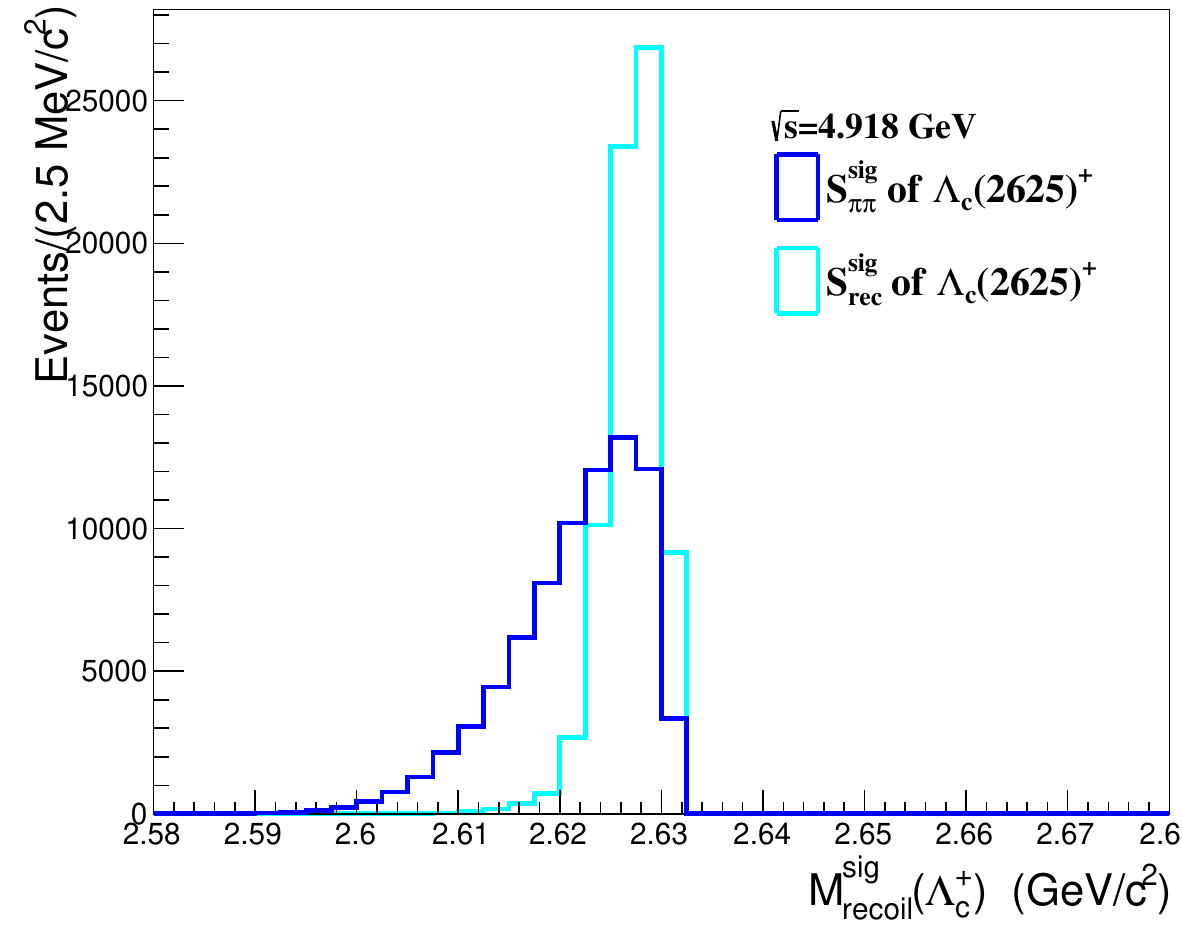}~\label{fig:Mrec-4915}}
            \subfigure{\includegraphics[width=0.42\textwidth]{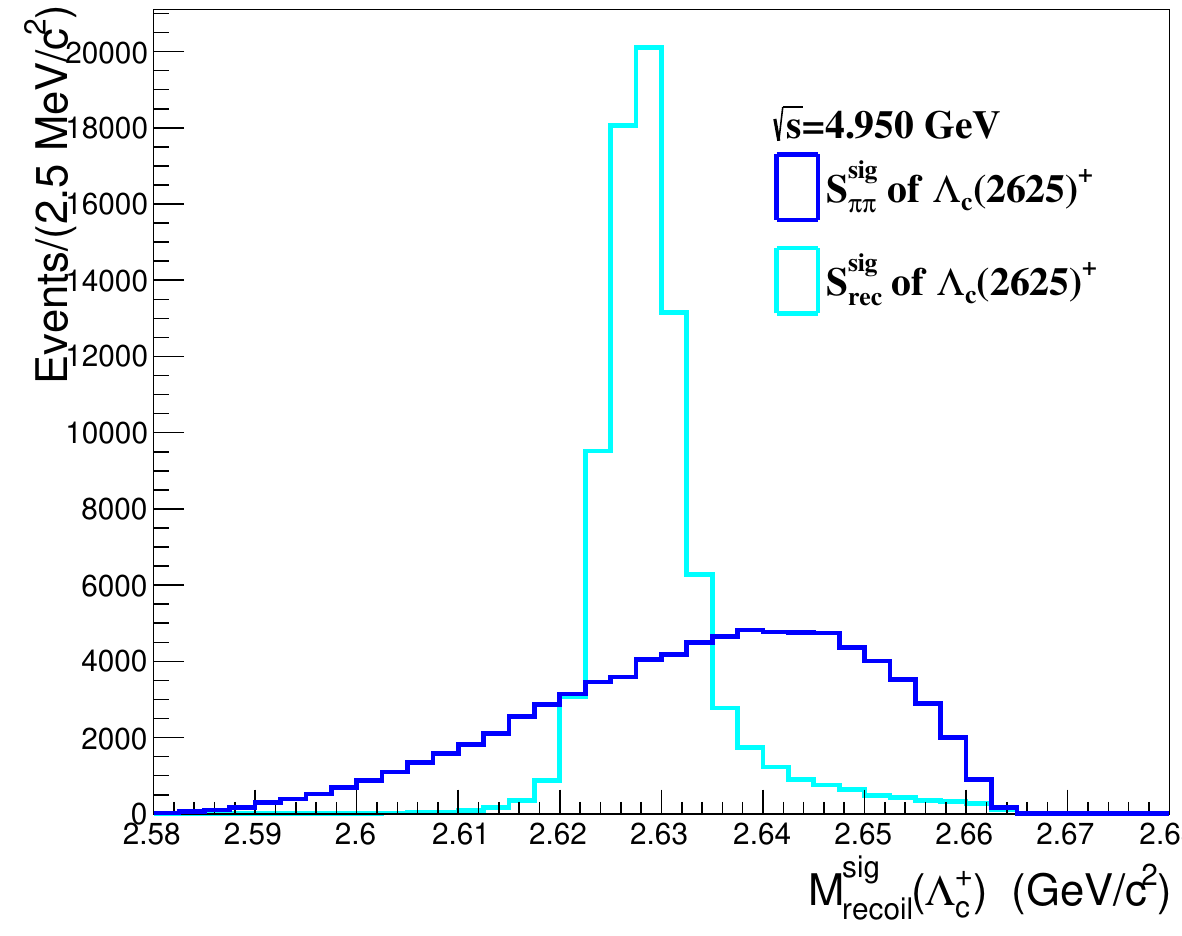}~\label{fig:Mrec-4946}}
	\end{center}
    \caption{The one dimensional distributions (a) and (b) \MLcpipi{} (c) and (d) \MrecLc{} of signal MC $S_{\pi\pi}^{\rm sig}$ and $S_{\rm rec}^{\rm sig}$ for $\sqrt{s} = 4.918$ (left) and 4.950~GeV (right). }
        \label{fig:MC1D}
\end{figure}
\begin{figure}[!htp]
    \begin{center}
            \subfigure{\includegraphics[width=0.42\textwidth]{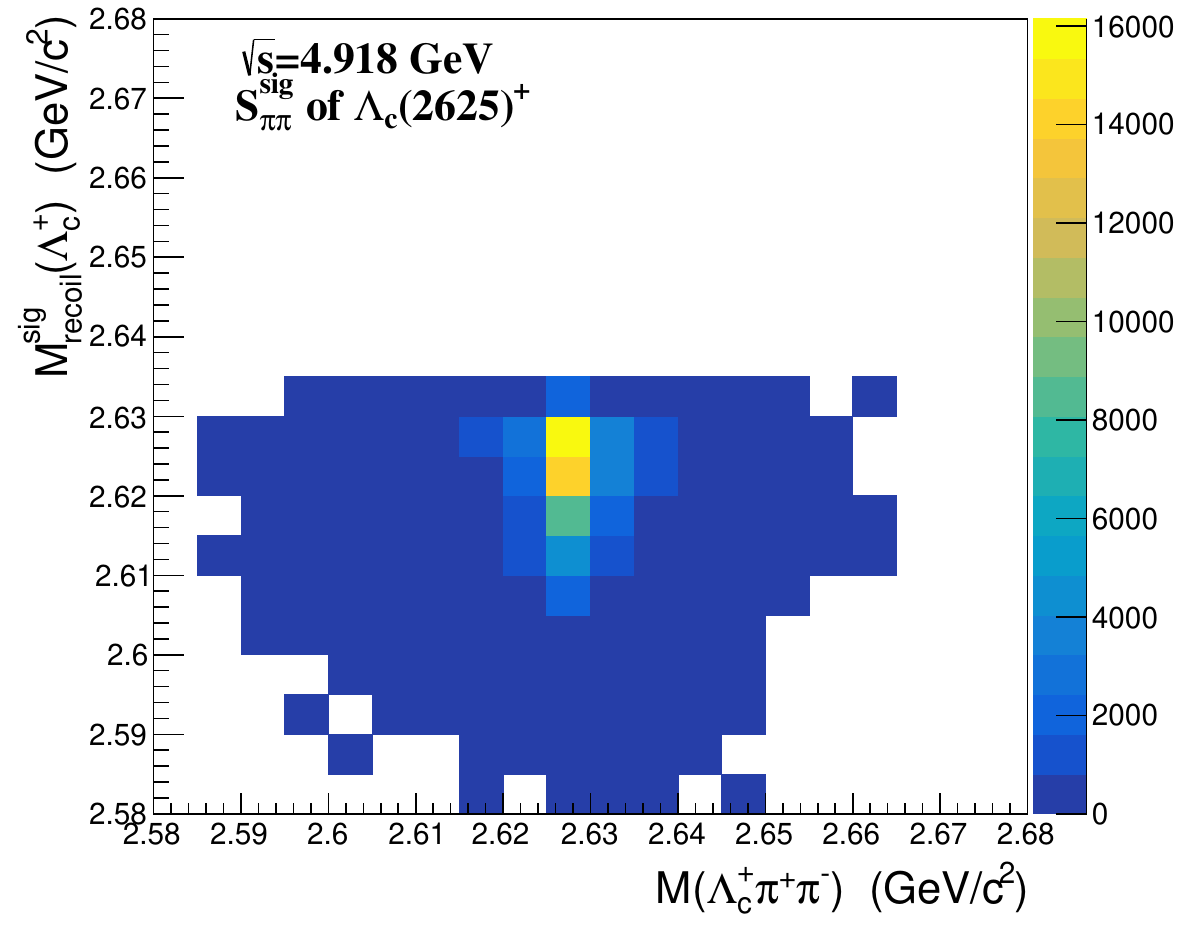}~\label{fig:Spipi-4915}} 
            \subfigure{\includegraphics[width=0.42\textwidth]{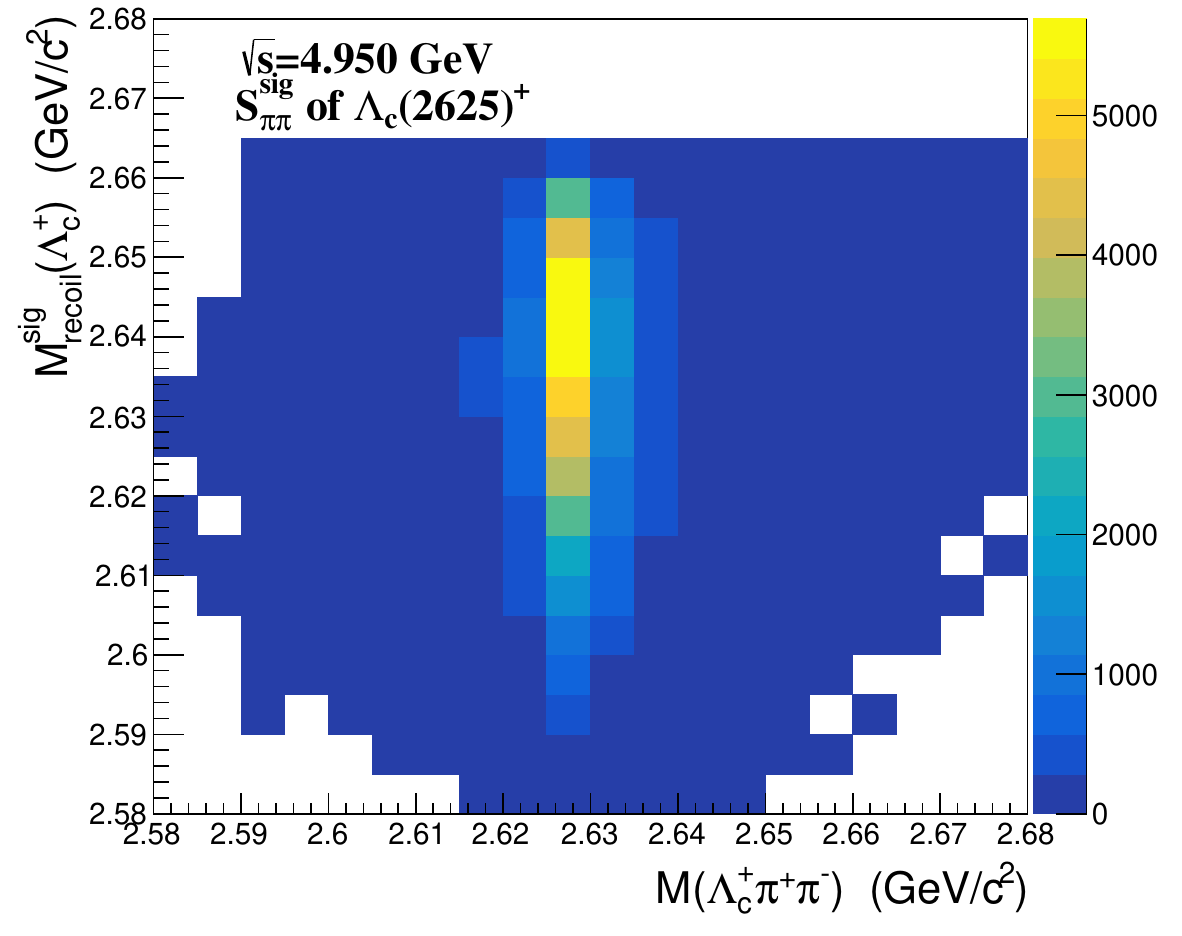}~\label{fig:Spipi-4946}}  
            
            \subfigure{\includegraphics[width=0.42\textwidth]{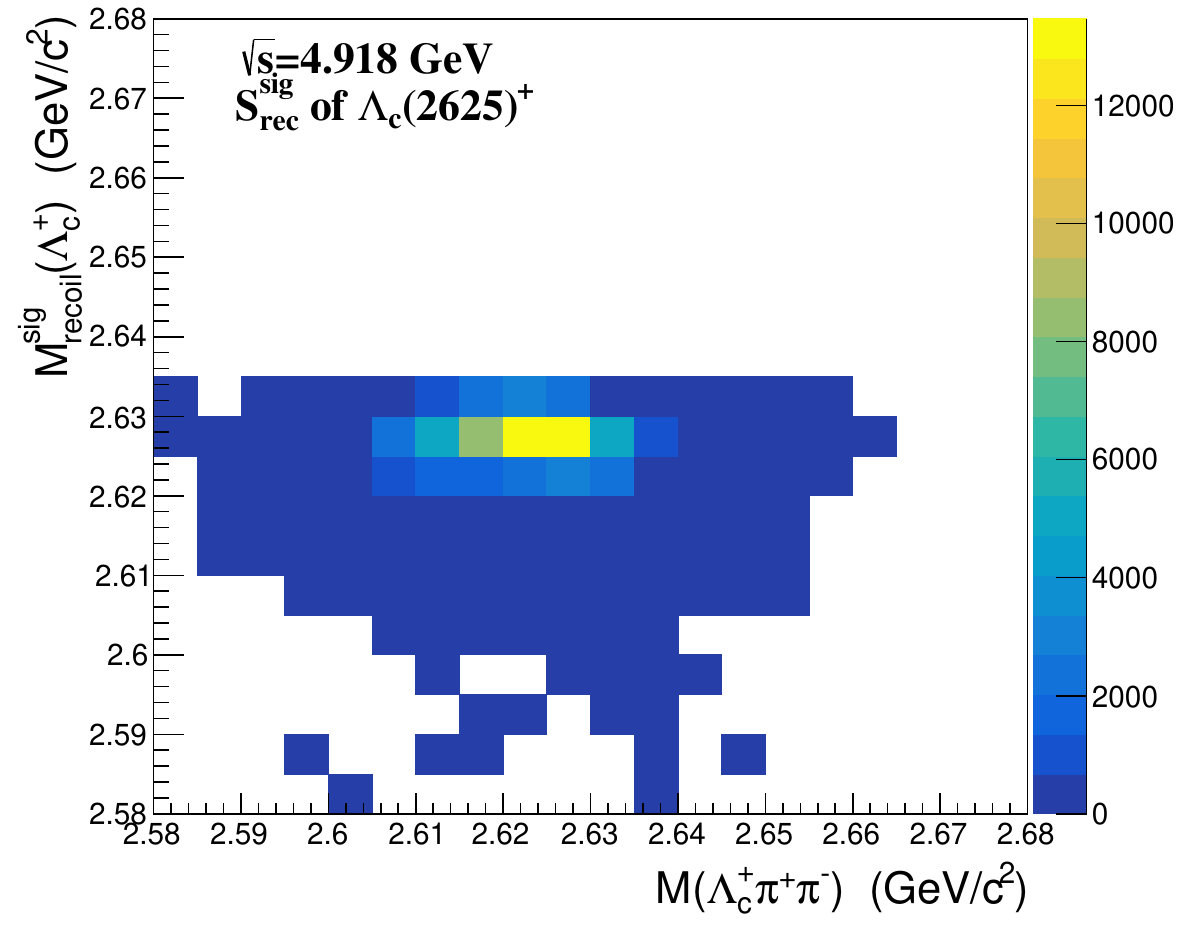}~\label{fig:Srec-4915}}
            \subfigure{\includegraphics[width=0.42\textwidth]{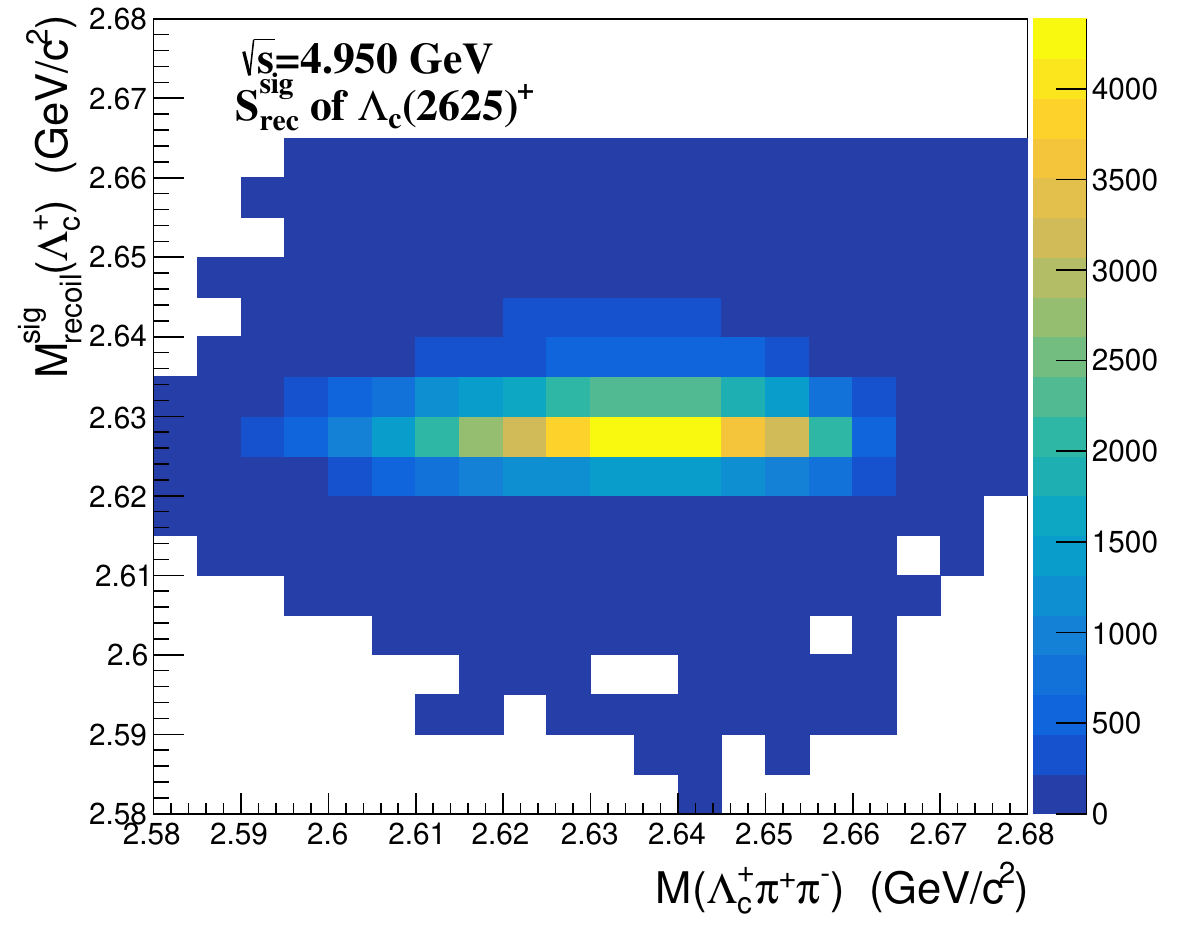}~\label{fig:Srec-4946}}
	\end{center}
    \caption{The two dimensional distributions \MLcpipi{} versus \MrecLc{} of signal MC $S_{\pi\pi}^{\rm sig}$ and $S_{\rm rec}^{\rm sig}$ for $\sqrt{s} = 4.918$ (left) and 4.950~GeV (right). }
        \label{fig:MC1D}
\end{figure}
\begin{figure}[!htp]
    \begin{center}
        \includegraphics[width=0.42\textwidth]{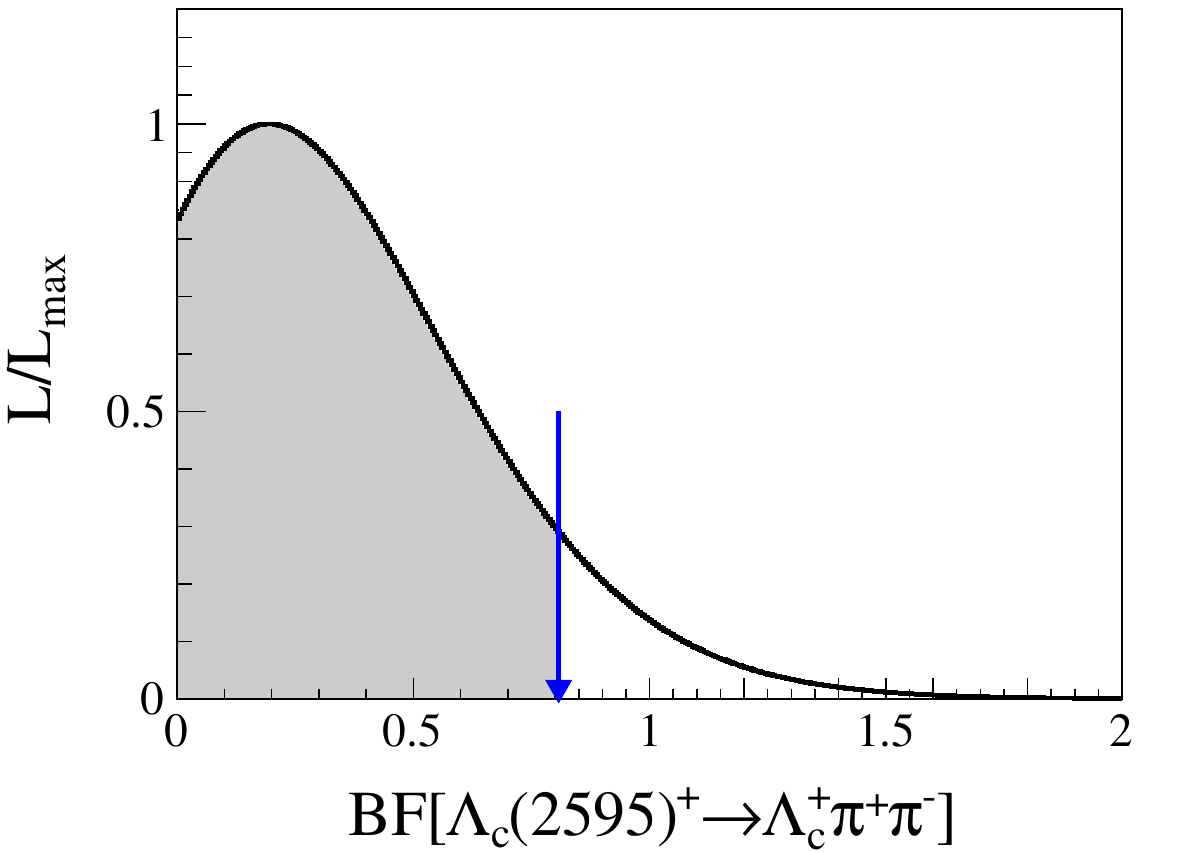}
    \end{center}
    \caption{
      Likelihood distributions over the BF of \LamCstarA{}. The black solid curve is the scan result with systematic uncertainties. The blue arrow indicates the upper limit of the BF at 90\% C.L.}
     \label{fig:single-tag-470}
\end{figure}

%% file: authorlist_2023-08-14.tex
\author{
   \begin{small}
     \begin{center}
M.~Ablikim$^{1}$, M.~N.~Achasov$^{4,b}$, P.~Adlarson$^{75}$, X.~C.~Ai$^{80}$, R.~Aliberti$^{35}$, A.~Amoroso$^{74A,74C}$, M.~R.~An$^{39}$, Q.~An$^{71,58}$, Y.~Bai$^{57}$, O.~Bakina$^{36}$
% ...
\\
\vspace{0.2cm}
(BESIII Collaboration)\\
\vspace{0.2cm} {\it
$^{1}$ Institute of High Energy Physics, Beijing 100049, People's Republic of China\\
% ...
$^{4}$ Budker Institute of Nuclear Physics SB RAS (BINP), Novosibirsk 630090, Russia\\
% ...
$^{35}$ Johannes Gutenberg University of Mainz, Johann-Joachim-Becher-Weg 45, D-55099 Mainz, Germany\\
$^{36}$ Joint Institute for Nuclear Research, 141980 Dubna, Moscow region, Russia\\
% ...
$^{39}$ Liaoning Normal University, Dalian 116029, People's Republic of China\\
% ...
$^{57}$ Southeast University, Nanjing 211100, People's Republic of China\\
$^{58}$ State Key Laboratory of Particle Detection and Electronics, Beijing 100049, Hefei 230026, People's Republic of China\\
% ...
$^{71}$ University of Science and Technology of China, Hefei 230026, People's Republic of China\\
% ...
$^{74}$ University of Turin and INFN, (A)University of Turin, I-10125, Turin, Italy; (B)University of Eastern Piedmont, I-15121, Alessandria, Italy; (C)INFN, I-10125, Turin, Italy\\
$^{75}$ Uppsala University, Box 516, SE-75120 Uppsala, Sweden\\
% ...
$^{80}$ Zhengzhou University, Zhengzhou 450001, People's Republic of China\\
\vspace{0.2cm}
% ...
$^{b}$ Also at the Novosibirsk State University, Novosibirsk, 630090, Russia\\
% ...
}
\end{center}
\vspace{0.4cm}
\end{small}
}

%% file: LcPiPi-draft-CL-FH.bbl
\begin{thebibliography}{99} 


\bibitem{Cheng:2021qpd}
H. Y. Cheng, 
\href{https://doi.org/10.1016/j.cjph.2022.06.021}{Chin. J. Phys. {\bf 78}, 324 (2022).}

\bibitem{PDG:2022}
R. L. Workman {\it et al.} (Particle Data Group), 
\href{https://doi.org/10.1093/ptep/ptac097}{Prog. Theor. Exp. Phys. {\bf 2022}, 083C01 (2022)}

\bibitem{Edwards:1995prl}
K.W. Edwards \textit{et al.} (CLEO Collaboration),
\href{https://doi.org/10.1103/PhysRevLett.74.3331}{Phys. Rev. Lett. {\bf 74}, 3331 (1995).}

\bibitem{Albrecht:1997plb}
H. Albrecht \textit{et al.} (ARGUS Collaboration),
\href{https://doi.org/10.1016/S0370-2693(97)00503-0}{Phys. Lett. \textbf{B} {\bf 402}, 207 (1997).}

\bibitem{PhysRevD.107.032008}
D. Wang \textit{et al.} (Belle Collaboration),
\href{https://link.aps.org/doi/10.1103/PhysRevD.107.032008}{Phys. Rev. \textbf{D} {\bf 107}, 032008 (2023).}

\bibitem{PhysRevD.75.014006} H. Y. Cheng and C. K. C, \href{https://link.aps.org/doi/10.1103/PhysRevD.75.014006}{Phys. Rev. \textbf{D} \textbf{75}, 014006 (2007).}

\bibitem{PhysRevD.92.074014} H. Y. Cheng and C. K. C, \href{https://link.aps.org/doi/10.1103/PhysRevD.92.074014}{Phys. Rev. \textbf{D} \textbf{92}, 074014 (2015).}

\bibitem{PhysRevD.46.1148} T. M. Yan, H. Y. Cheng, C. Y. Cheung, G. L. Lin, Y. C. Lin, and H. L. Yu, \href{https://doi.org/10.1103/PhysRevD.46.1148}{Phys. Rev. \textbf{D} \textbf{46}, 1148 (1992).}

\bibitem{PhysRevD.56.5483} D. Pirjol and T. M. Yan, \href{https://doi.org/10.1103/PhysRevD.56.5483}{Phys. Rev. \textbf{D} \textbf{56}, 5483 (1997).}

\bibitem{PhysRevD.67.074033} E. B. Andrew and F. F. Adam and P. Dan and M. Y. John, \href{https://doi.org/10.1103/PhysRevD.67.074033}{Phys. Rev. \textbf{D} \textbf{67}, 074033 (2003).}

\bibitem{Frabetti:1994prl}
P. L. Frabetti \textit{et al.} (E687 Collaboration),
\href{https://doi.org/10.1103/PhysRevLett.72.961}{Phys. Rev. Lett. {\bf 72}, 961 (1994).}

\bibitem{Frabetti:1996plb}
P. L. Frabetti \textit{et al.} (E687 Collaboration),
\href{https://doi.org/10.1016/0370-2693(95)01458-6}{Phys. Lett. \textbf{B} {\bf 365}, 461 (1996).}

\bibitem{Albrecht:1993plb}
H. Albrecht \textit{et al.} (ARGUS Collaboration),
\href{https://doi.org/10.1016/0370-2693(93)91598-H}{Phys. Lett. \textbf{B} {\bf 317}, 227 (1993).}

\bibitem{Aaltonen:2011prb}
T. Aaltonen \textit{et al.} (CDF Collaboration),
\href{https://doi.org/10.1103/PhysRevD.84.012003}{Phys. Rev. \textbf{D} {\bf 84}, 012003 (2011).}

\bibitem{PhysRevD.95.114018}
A. J. Arifi, H. Nagahiro, and A. Hosaka,
\href{https://doi.org/10.1103/PhysRevD.95.114018}{Phys. Rev. \textbf{D} {\bf 95}, 114018 (2017).}

\bibitem{PhysRevD.92.014036}
J. X. Lu, Y. Zhou, H. X. Chen, J. J. Xie, and L. S. Geng,
\href{https://doi.org/10.1103/PhysRevD.92.014036}{Phys. Rev. \textbf{D} {\bf 92}, 014036 (2015).}

\bibitem{PhysRevD.101.014018}
J. Nieves and R. Pavao,
\href{https://doi.org/10.1103/PhysRevD.101.014018}{Phys. Rev. \textbf{D} {\bf 101}, 014018 (2020).}

\bibitem{EPJC.81.224}
Q. Zhang, X. H. Hu, B. R. He, and J. L. Ping,
\href{https://doi.org/10.1140/epjc/s10052-021-09017-8}{Eur. Phys. J. \textbf{C} {\bf 81}, 224 (2021).}

\bibitem{Moriya:2014prl}
K. Moriya \textit{et al.} (CLAS Collaboration),
\href{https://doi.org/10.1103/PhysRevLett.112.082004}{Phys. Rev. Lett. {\bf 112}, 082004 (2014).}

\bibitem{Kamano:2015prc}
H. Kamano, S. X. Nakamura, T.-S. H. Lee, and T. Sato,
\href{https://doi.org/10.1103/PhysRevC.92.025205}{Phys. Rev. \textbf{C} {\bf 92}, 025205 (2015).}

\bibitem{Baician:2023}
B.-C. Ke, J. Koponen, H. B. Li, and Y. H. Zheng,
\href{https://doi.org/10.1146/annurev-nucl-110222-044046}{ANNU. REV. NUCL. PART. S. {\bf 73}, 285 (2023).}

\bibitem{BESIII:Lumi} M. Ablikim \textit{et al.} (BESIII Collaboration), \href{https://dx.doi.org/10.1088/1674-1137/ac84cc}{Chin. Phys. \textbf{C} \textbf{46}, 113003 (2022).}

\bibitem{Ablikim:2009aa} M. Ablikim \textit{et al.} (BESIII Collaboration), \href{https://dx.doi.org/10.1016/j.nima.2009.12.050}{Nucl. Instrum. Meth. \textbf{A} \textbf{614}, 345 (2010).}

\bibitem{Agostinelli:2002hh} S. Agostinelli \textit{et al.} (GEANT4 Collaboration), \href{https://dx.doi.org/10.1016/S0168-9002(03)01368-8}{Nucl. Instrum. Meth. \textbf{A} \textbf{506}, 250 (2003).}

\bibitem{Kaixuan:2022}
K. X. Huang \textit{et al.}, 
\href{https://doi.org/10.1007/s41365-022-01133-8}{Nucl. Sci. Tech. { \bf 33}, 142 (2022).}

\bibitem{Lange:2001uf} D. J. Lange, \href{https://dx.doi.org/10.1016/S0168-9002(01)00089-4}{Nucl. Instrum. Meth. \textbf{A} \textbf{462}, 152 (2001).}

\bibitem{Ping:2008zz} R. G. Ping, \href{https://dx.doi.org/10.1088/1674-1137/32/8/001}{Chin. Phys. \textbf{C} \textbf{32}, 599 (2008).}

\bibitem{Chen:2000tv} J. C. Chen, G. S. Huang, X. R. Qi, D. H. Zhang, and
Y. S. Zhu, \href{https://dx.doi.org/10.1103/PhysRevD.62.034003}{Phys. Rev. \textbf{D} \textbf{62}, 034003 (2000).}

\bibitem{PhysRevLett.31.061301} 
Y. L. Yang, R. G. Ping, and H. Chen, \href{https://doi.org/10.1088/0256-307X/31/6/061301}{Phys. Rev. Lett. {\bf 31}, 061301 (2014).}

\bibitem{Richter-Was:1992hxq} E. R. Was, \href{https://dx.doi.org/10.1016/0370-2693(93)90062-M}{Phys. Lett. \textbf{B} \textbf{303}, 163 (1993).}

\bibitem{Junhua:2023} M. Ablikim \textit{et al.} (BESIII Collaboration), 
\href{https://arxiv.org/abs/2312.08414}{arXiv:2312.08414}

\bibitem{Jadach:2000ir} S. Jadach, B. F. L. Ward, and Z. Was, \href{https://dx.doi.org/10.1103/PhysRevD.63.113009}{Phys. Rev. \textbf{D} \textbf{63}, 113009 (2001).}

\bibitem{Supp:2023}
See Supplemental Material for additional details.

\bibitem{ARGUS:1990}  H. Albrecht \textit{et al.} (ARGUS Collaboration), \href{http://dx.doi.org/10.1016/0370-2693(90)91293-K}{Phys. Lett. \textbf{B} \textbf{241}, 278 (1990).}

\bibitem{stenson2006exact} K. Stenson, \href{https://arxiv.org/abs/physics/0605236} {arXiv physics/0605236}

\bibitem{Weiping:2023} M. Ablikim \textit{et al.} (BESIII Collaboration), \href{https://arxiv.org/abs/2307.07316}{arXiv 2307.07316 (2023).}

\end{thebibliography}
